\documentclass[acmsmall]{acmart}

\usepackage{array}
\usepackage{arydshln}
\usepackage{graphicx}
\newcommand{\heading}[1]{{\vspace{3pt}\noindent{\textbf{#1}}}}
\usepackage{multirow}
\newcommand{\fullcov}{\CIRCLE}
\newcommand{\partcov}{\LEFTcircle}
\newcommand{\nocov}{\Circle}
\usepackage{wasysym}
\AtBeginDocument{%
  }

\makeatletter
\AtBeginDocument{%
  \let\originalcopyrightpermission\@copyrightpermission
  \def\@copyrightpermission{%
    This research has received support from National Key R\&D Program of China (2023YFB2704700), National Natural Science Foundation of China (62472276), and Science and Technology Project of the State Grid Corporation of China (5700-202321603A-3-2-ZN).\par
    \originalcopyrightpermission}}
\makeatother

\newcolumntype{L}[1]{>{\raggedright\arraybackslash}p{#1}}
\newcolumntype{M}[1]{>{\centering\arraybackslash}p{#1}}
\newcommand{\surveyfig}[1]{%
  \includegraphics[width=\linewidth,height=0.36\textheight,keepaspectratio]{#1}}

\begin{document}

\title{Blockchain Attacks and Defenses: A Layered and Cross-Domain Survey}

\author{Junjie Hu}
\email{nakamoto@sjtu.edu.cn}
\orcid{0009-0008-8524-9070}
\author{Na Ruan}
\email{naruan@sjtu.edu.cn}
\orcid{0000-0002-7673-9843}
\affiliation{%
  \institution{School of Computer Science, Shanghai Jiao Tong University}
  \city{Shanghai}
  \country{China}
}

\renewcommand{\shortauthors}{Hu and Ruan}

\begin{abstract}
Blockchains have evolved from simple distributed ledgers into programmable platforms that process complex application logic and carry significant financial value. All modern Web3 systems share a common goal: providing secure, decentralized, and trustworthy execution in an increasingly interconnected environment. However, this evolution has shifted the attack surface from isolated infrastructure disruptions to programmable economic abuse and cross-domain exploits. In this article, we focus on the research of blockchain attacks and defenses. In particular, we categorize the threat landscape and corresponding mitigation strategies according to both a four-tier layered architecture (network, cryptographic, consensus, and application) and cross-domain trust boundaries. We seek to answer these important questions: How has the research in blockchain security evolved over the past decade, especially with the rise of decentralized finance (DeFi) and cross-chain interoperability? How do local security assumptions fail when protocols are composed, and what are the driving needs for Web3 security research in the future?
\end{abstract}

\ccsdesc[500]{Security and privacy~Distributed systems security}
\ccsdesc[300]{Security and privacy~Economics of security and privacy}

\keywords{Blockchain security, Web3 security, smart contracts, DeFi, consensus, MEV, cross-chain bridges}

\maketitle

\section{Introduction}
\label{sec:introduction}

\subsection{Background and Motivation}

Over the past decade, blockchains have evolved from single-asset ledgers \cite{web3sec001_sokresearchperspectiveschallengesbitcoin} into programmable settlement platforms. Unlike distributed databases focused on replication \cite{web3sec012_blockchainsvsdistributeddatabasesdichotomy}, smart-contract systems execute DeFi and Web3 workflows under open, irreversible, and financially valuable conditions. This coupling changes the threat model: a small implementation bug or economic misconfiguration can immediately become a large loss.

\begin{figure}[t]
  \centering
  \surveyfig{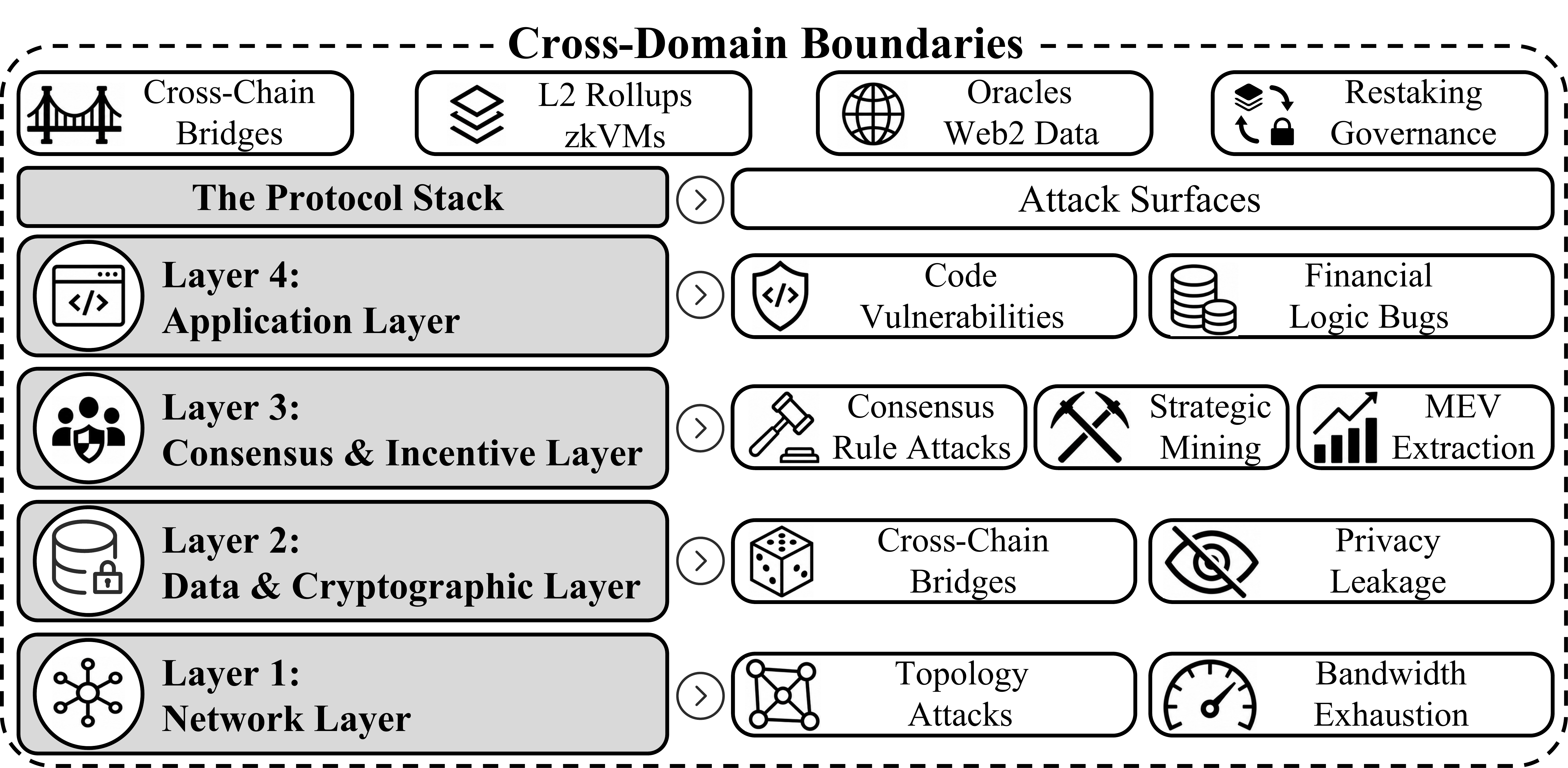}
  \caption{A layered and cross-domain taxonomy for blockchain attacks and defenses.}
  \Description{A layered taxonomy that organizes Web3 security into network, data and cryptographic, consensus and incentive, application, and cross-domain components.}
  \label{fig:taxonomy}
\end{figure}

\begin{figure}[t]
  \centering
  \surveyfig{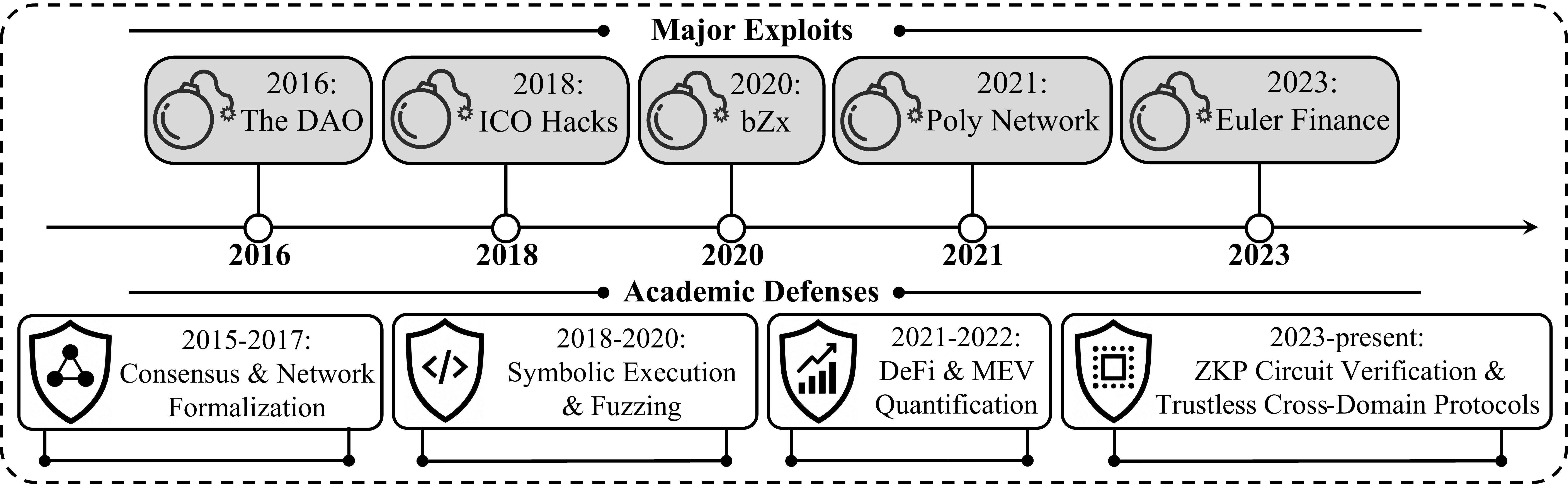}
  \caption{Evolution of dominant Web3 security threats.}
  \Description{A timeline showing the evolution from P2P and consensus attacks to smart-contract vulnerabilities and modern cross-domain DeFi threats.}
  \label{fig:timeline}
\end{figure}

The locus of blockchain security has shifted from infrastructure disruption to programmable economic abuse (Figure~\ref{fig:timeline}). Early attacks targeted P2P networks and consensus bounds; later work focused on smart-contract bytecode flaws. Modern attacks increasingly exploit assumption mismatches---the gap between a protocol's local guarantee and how other systems consume it. Recent work on PoS incentives, fee rules, MEV, validator deanonymization, and zkEVM soundness shows that these risks now interact within the same adversarial economy \cite{web3sec201_auspextransactionfee,web3sec202_availableattestation,web3sec203_bunnyfinderethereumconsensus,web3sec205_ethereumvalidatordeanonymizing,web3sec210_polygonzkevmfreever,web3sec212_lightdarknessmevbot}. Figure~\ref{fig:taxonomy} summarizes the resulting taxonomy across protocol assumptions, execution incentives, and cross-domain trust boundaries.

\subsection{Limitations of Existing Surveys}

Existing blockchain-security surveys and SoK papers leave three gaps for modern Web3:

\begin{itemize}
    \item \textbf{Outdated scope.} Foundational surveys emphasize Bitcoin-era network and consensus mechanisms \cite{web3sec001_sokresearchperspectiveschallengesbitcoin,web3sec013_securityprivacyblockchain}, leaving DeFi, oracles, and Layer-2 failures underdeveloped.
    \item \textbf{Siloed analysis.} Empirical studies often isolate Solidity bugs or scanners \cite{web3sec014_blockchainsmartcontractsecuritythreats,web3sec015_empiricalstudyblockchainsystemvulnerabilities}, while modern losses combine network assumptions, consensus incentives, MEV, and application logic.
    \item \textbf{Uneven evidence.} Some surveys mix mature security research with lightly validated preprints, making it harder to identify the research frontier.
\end{itemize}

Prior taxonomies usually classify by component, vulnerability class, or analysis technique. Specialized SoKs cover DeFi, MEV, bridges, oracles, restaking, DAOs, AI-aided analysis, and SNARK vulnerabilities \cite{web3sec174_sokdecentralizedfinancedefiattacks,web3sec177_sokpreventingtransactionreorderingmanipulations,web3sec244_sokmevcountermeasures,web3sec196_soksecuritycrosschainbridges,web3sec245_sokoraclesgroundtruth,web3sec246_sokliquidstakingrestaking,web3sec247_sokdefifundamentalstaxonomyrisks,web3sec248_sokattacksdaos,web3sec198_sokaipoweredsecurityanalysis,web3sec231_soksnarkvulnerabilities}. Table~\ref{tab:survey-surveys} positions this survey against representative SoKs. Our key difference is to ask how assumptions move across layers: a bridge exploit may start as weak validation, appear as accounting error on another chain, and end as a liquidity crisis; an MEV attack may start with transaction visibility, exploit consensus ordering, and drain an application with no conventional code bug.

\begin{table*}[t]
\centering
\caption{Survey-of-surveys comparison and positioning of this work.}
\label{tab:survey-surveys}
\scriptsize
\setlength{\tabcolsep}{3.2pt}
\renewcommand{\arraystretch}{1.03}
\setlength{\dashlinedash}{0.65pt}
\setlength{\dashlinegap}{0.8pt}
\setlength{\arrayrulewidth}{0.35pt}
\resizebox{\textwidth}{!}{%
\begin{tabular}{|c|c|c|c|c|c|c|c|}
\hline
\textbf{Existing survey / SoK} &
\textbf{Scope} &
\textbf{Taxonomy axis} &
\textbf{MEV} &
\textbf{Bridges / rollups} &
\textbf{Oracle / intent / restaking} &
\textbf{Defense trade-offs} &
\textbf{Main limitation for this survey} \\
\hline\hline
Bitcoin and cryptocurrency SoK \cite{web3sec001_sokresearchperspectiveschallengesbitcoin} &
Bitcoin-era foundations &
Protocol primitives &
\nocov & \nocov & \nocov & \partcov &
Pre-Web3 scope \\
\hdashline
Blockchain security and privacy survey \cite{web3sec013_securityprivacyblockchain} &
General blockchain security &
Security properties &
\nocov & \nocov & \nocov & \partcov &
Limited DeFi/modular coverage \\
\hdashline
Smart-contract security survey \cite{web3sec014_blockchainsmartcontractsecuritythreats} &
Smart contracts &
Vulnerability classes &
\partcov & \nocov & \partcov & \partcov &
Application-centric \\
\hdashline
Empirical vulnerability study \cite{web3sec015_empiricalstudyblockchainsystemvulnerabilities} &
System vulnerabilities &
Modules and bug types &
\nocov & \partcov & \nocov & \nocov &
No unified trust-boundary model \\
\hdashline
DeFi attacks SoK \cite{web3sec174_sokdecentralizedfinancedefiattacks} &
DeFi exploits &
Financial attack mechanisms &
\partcov & \partcov & \partcov & \partcov &
Limited lower-layer linkage \\
\hdashline
DeFi fundamentals SoK \cite{web3sec247_sokdefifundamentalstaxonomyrisks} &
DeFi protocols and risks &
Protocol stack and risk types &
\partcov & \partcov & \partcov & \partcov &
Broad but DeFi-bounded \\
\hdashline
Transaction reordering SoK \cite{web3sec177_sokpreventingtransactionreorderingmanipulations} &
Ordering manipulation &
MEV and ordering defenses &
\fullcov & \nocov & \nocov & \fullcov &
Ordering-specific \\
\hdashline
MEV countermeasures SoK \cite{web3sec244_sokmevcountermeasures} &
MEV mitigation &
Countermeasure design &
\fullcov & \partcov & \nocov & \fullcov &
MEV-defense specific \\
\hdashline
Cross-chain bridge SoK \cite{web3sec196_soksecuritycrosschainbridges} &
Cross-chain bridges &
Bridge trust models &
\nocov & \partcov & \partcov & \partcov &
Bridge-specific \\
\hdashline
Oracle SoK \cite{web3sec245_sokoraclesgroundtruth} &
Blockchain oracles &
Data-source trust models &
\partcov & \partcov & \fullcov & \partcov &
Oracle-specific \\
\hdashline
Liquid staking/restaking SoK \cite{web3sec246_sokliquidstakingrestaking} &
Staking derivatives &
Collateral and operator risks &
\nocov & \partcov & \fullcov & \partcov &
Staking-market specific \\
\hdashline
DAO attacks SoK \cite{web3sec248_sokattacksdaos} &
DAO governance &
Governance attack patterns &
\nocov & \nocov & \partcov & \partcov &
Governance-specific \\
\hdashline
AI-powered smart-contract SoK \cite{web3sec198_sokaipoweredsecurityanalysis} &
AI-aided analysis &
Tool capability &
\nocov & \nocov & \nocov & \partcov &
Tool-centric \\
\hdashline
SNARK vulnerabilities SoK \cite{web3sec231_soksnarkvulnerabilities} &
SNARK security &
Proof-system bugs &
\nocov & \partcov & \nocov & \partcov &
ZK-specific \\
\hdashline
\textbf{This survey} &
\textbf{Web3 attack/defense stack} &
\textbf{Assumption propagation} &
\fullcov & \fullcov & \fullcov & \fullcov &
\textbf{Unified cross-layer model} \\
\hline
\end{tabular}%
}
\vspace{2pt}
\begin{flushleft}
\scriptsize
\fullcov = substantial coverage; \partcov = scope-limited coverage; \nocov = absent coverage.
\end{flushleft}
\end{table*}

\subsection{Methodology and Paper Selection}

\begin{figure*}[t]
  \centering
  \includegraphics[width=0.92\textwidth]{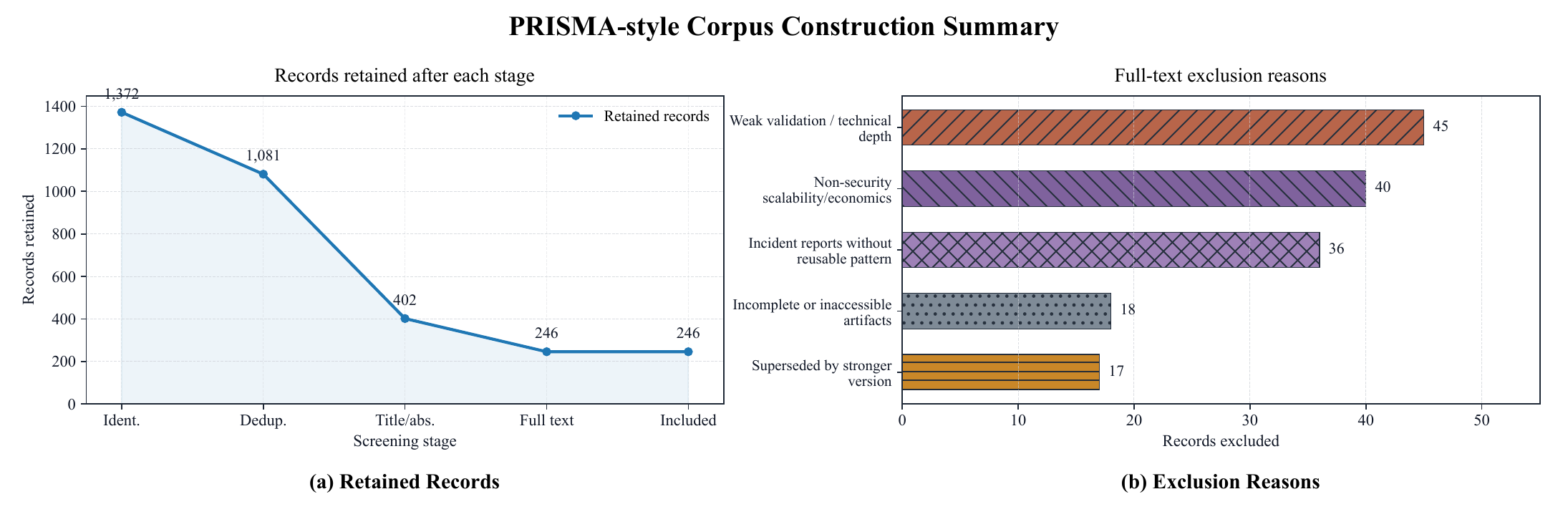}
  \caption{PRISMA-style corpus construction flow. Counts provide an audit trail for reproducibility. Public incident reports used solely for empirical case tables are excluded from the academic retained-study count.}
  \Description{A PRISMA-style flow diagram showing identified records, records retained after deduplication, records retained after title and abstract screening, records retained after full-text eligibility assessment, and studies included in synthesis. The figure also shows retained-record bars and full-text exclusion reasons.}
  \label{fig:prisma-flow}
\end{figure*}

We conducted a systematic review following PRISMA guidelines (Figure~\ref{fig:prisma-flow}). The screening process uses a \emph{problem-driven inclusion rule}: a study is retained only if it changes how a blockchain system is modeled, exploited, or defended. We exclude purely descriptive incident reports without reusable attack patterns, while including scalability and interoperability protocols when their primitives introduce new adversarial surfaces.

The corpus, updated through June 2026, was built from major digital libraries, official proceedings, arXiv, and IACR ePrint. Search queries paired Web3 components---smart contracts, DeFi, cross-domain infrastructure, MEV, restaking, and intents---with adversarial-analysis terms, then expanded by backward and forward snowballing from seminal SoKs, top-tier security proceedings, and highly cited systems papers.

During full-text screening, we separated three roles that are often mixed in survey corpora. First, a paper may define a new system abstraction, such as a consensus protocol, bridge design, rollup architecture, or proof mechanism. Second, it may expose an attack pattern that generalizes beyond one deployment. Third, it may provide a defense or measurement method whose assumptions can be compared across systems. We retained papers that clearly served at least one of these roles. We excluded reports that only recount losses or operational compromise without a reusable security pattern, and we treated recent incident reports only as empirical anchors for case tables. This choice keeps the corpus focused on transferable knowledge rather than raw event volume.

\begin{figure*}[t]
  \centering
  \includegraphics[width=\textwidth]{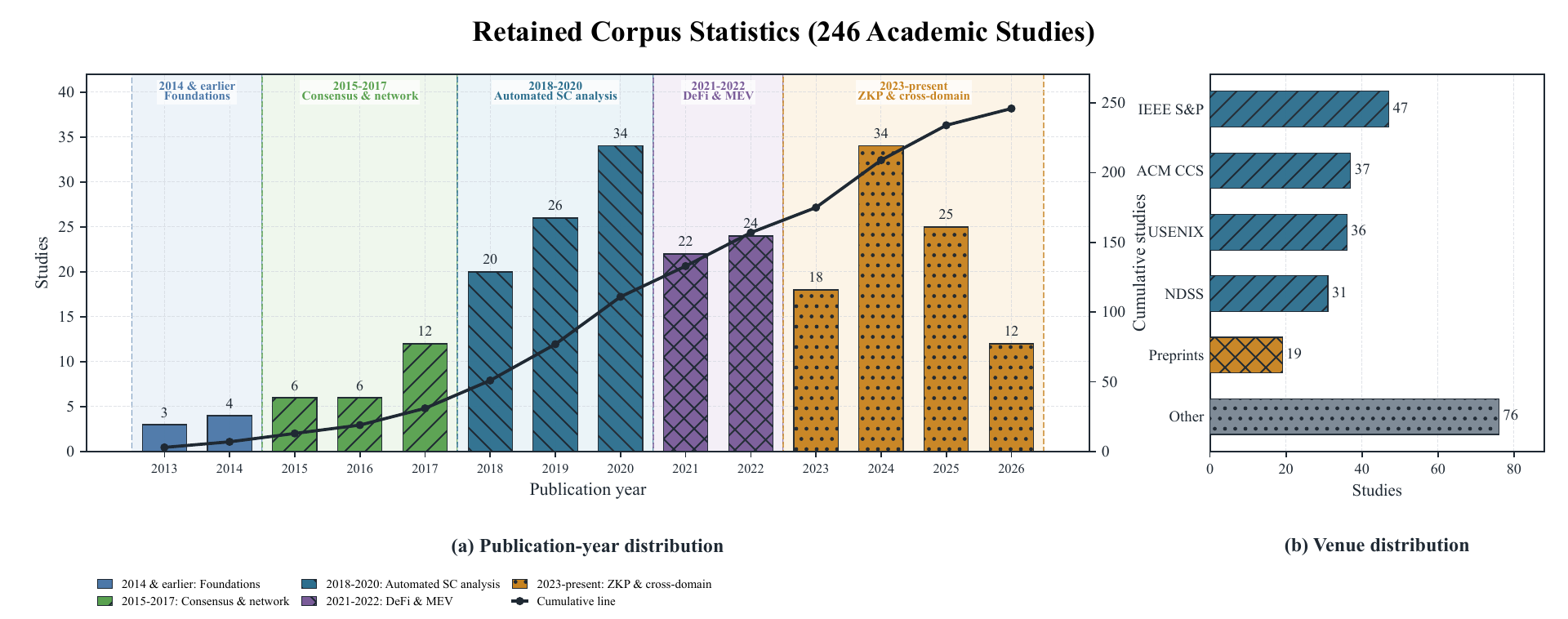}
  \caption{Retained-corpus statistics. The left panel reports publication-year distribution and cumulative growth; the right panel reports venue distribution. Counts provide an audit trail for reproducibility and exclude public incident reports used solely for empirical case tables.}
  \Description{A two-panel retained-corpus statistics figure. The left panel occupies about three quarters of the width and shows annual retained academic studies from 2013 to 2026 with a cumulative growth line. The right panel occupies about one quarter of the width and shows venue counts for IEEE S and P, ACM CCS, USENIX Security, NDSS, preprints, and other venues.}
  \label{fig:corpus-statistics}
\end{figure*}

As detailed in Table~\ref{tab:literature-scope}, the final corpus comprises 246 high-quality studies. Figure~\ref{fig:corpus-statistics} summarizes both publication years and venues: the corpus is concentrated in the \textit{Big Four} security conferences, while also extending into software-engineering, distributed-systems, cryptography venues, and preprint archives. Peer-reviewed papers are dated by official proceedings year, while preprints are dated by first public release to avoid conflating archival publication with preprint availability.

\begin{table*}[t]
\centering
\caption{Literature corpus and analytical scope.}
\label{tab:literature-scope}
\scriptsize
\setlength{\tabcolsep}{4.6pt}
\renewcommand{\arraystretch}{1.05}
\setlength{\dashlinedash}{0.65pt}
\setlength{\dashlinegap}{0.95pt}
\setlength{\arrayrulewidth}{0.35pt}
\resizebox{\textwidth}{!}{%
\begin{tabular}{|c|c|c|c|}
\hline
\textbf{Scope} &
\textbf{Representative venues} &
\textbf{Retained studies} &
\textbf{Main analytical focus} \\
\hline\hline
Foundations and surveys &
IEEE S\&P, CACM, ACM CSUR, SIGMOD, SOSP &
20 &
Ledger models, blockchain/database comparison, core system abstractions, prior SoKs \\
\hdashline
Network and consensus &
IEEE S\&P, ACM CCS, NDSS, FC, EuroSys &
99 &
P2P topology, routing attacks, PoW/PoS safety, BFT protocols, incentives, MEV \\
\hdashline
Cryptography and privacy &
IEEE S\&P, ACM CCS, EUROCRYPT, PETS &
38 &
Anonymity, zero-knowledge proofs, threshold randomness, post-quantum confidential transactions \\
\hdashline
Smart contracts and DeFi &
ACM CCS, USENIX Security, ICSE, ISSTA, NDSS &
73 &
Static analysis, fuzzing, formal verification, flash loans, economic exploits \\
\hdashline
Cross-domain systems &
ACM CCS, IEEE S\&P, NDSS, USENIX Security &
16 &
Bridges, rollups, payment channels, oracles, cross-layer attack composition \\
\hline
\end{tabular}%
}
\vspace{2pt}
\begin{flushleft}
\scriptsize
Retained-study counts reflect mutually exclusive primary-topic assignments (totaling 246). Papers relevant to multiple layers are assigned to the category that best matches their primary technical contribution. 
\end{flushleft}
\end{table*}

\subsection{Research Questions and Scope}

This survey asks four questions, summarized in Table~\ref{tab:rq}. Where do vulnerabilities originate in each architectural layer? Do defenses fix the failed assumption or only block a known exploit? How do local failures cascade across layers and trust domains? Which abstractions are still missing for the next generation of Web3 systems?

\begin{table*}[t]
\centering
\caption{Research questions and survey coverage.}
\label{tab:rq}
\setlength{\tabcolsep}{4.6pt}
\renewcommand{\arraystretch}{1.08}
\setlength{\dashlinedash}{0.65pt}
\setlength{\dashlinegap}{0.8pt}
\setlength{\arrayrulewidth}{0.15pt}
\resizebox{\textwidth}{!}{%
\begin{tabular}{|c|c|c|}
\hline
\textbf{Question} &
\textbf{Focus} &
\textbf{Coverage} \\
\hline\hline
RQ1 &
What are the dominant attack surfaces in each architectural layer? &
Sections~\ref{sec:network}--\ref{sec:application} \\
\hdashline
RQ2 &
How do defenses address root causes, assumptions, and trade-offs? &
Sections~\ref{sec:network}--\ref{sec:future} \\
\hdashline
RQ3 &
How do local assumptions fail when protocols are composed? &
Section~\ref{sec:crossdomain} \\
\hdashline
RQ4 &
Which problems remain open for the next generation of Web3 systems? &
Section~\ref{sec:future} \\
\hline
\end{tabular}%
}
\vspace{2pt}
\begin{flushleft}
\scriptsize
The questions connect the layer-by-layer taxonomy to the cross-domain analysis and future research agenda.
\end{flushleft}
\end{table*}

We define \emph{blockchain security} through protocol and system design. A study is in scope if it explains how a decentralized system can lose assets, corrupt state, leak privacy, or suffer downtime. Routine exchange hacks, wallet phishing, and custodial breaches are excluded unless they expose a reusable design flaw. Our organizing question is simple: \emph{what assumption failed, and who depended on it?} We therefore judge a scanner, bridge verifier, or MEV mitigation by the invariant it preserves, the incentives it changes, and the deployment conditions under which it fails.

This scope also shapes how we use incidents. A large loss is not automatically a research contribution, and a small proof-of-concept attack can be highly important if it changes the model. Conversely, a public incident can be useful when it illustrates how several assumptions failed at once. We therefore use incidents to stress-test the taxonomy, not to rank protocols by loss size. This keeps the survey aligned with its main question: whether a defense would still hold when the same assumption is imported by another layer, another market, or another trust domain.

\subsection{Contributions}

This survey is organized as a coding and comparison exercise rather than a catalog of incidents. The main contributions are:

\begin{itemize}
  \item \textbf{Layered attack coding.} Each attack is encoded by origin layer, entry condition, violated invariant, affected asset or service, and representative defense. Table~\ref{tab:consensus-threats} applies this coding to consensus and MEV cases, while Table~\ref{tab:defi-incidents} maps major application and bridge incidents to reusable root causes.
  \item \textbf{Defense residual assumptions.} For scanners, privacy mechanisms, consensus changes, MEV mitigations, and bridge or rollup defenses, the survey separates the property checked by the defense from the deployment assumption that remains. This distinction is used repeatedly in the layer sections, for example when comparing static analysis with exploitability evidence and ZK bridge verification with proof-system implementation risk.
  \item \textbf{Cross-domain claim schema.} Bridges, rollups, oracles, intents, and restaking are compared with the same fields: imported claim, source trust model, destination consequence, timing window, and failure containment. Table~\ref{tab:crossdomain-failures} applies the schema to validation asymmetry, finality mismatch, oracle-source correlation, data-availability gaps, and shared-security overload.
  \item \textbf{Timing-based evaluation.} The survey records whether detection, response, and settlement occur before the loss becomes irreversible. This criterion is used to distinguish forensic tools from preventive monitors, fraud proofs that work under normal fees from those that fail under congestion, and intent receipts that bind solver behavior before cross-chain settlement from those that only explain failure afterward.
\end{itemize}

\subsection{Organization}

The rest of the survey follows the same logic. We first define the layered architecture and threat model. We then analyze where failures originate and move upward through the stack. The later sections focus on composition: how cross-domain systems carry assumptions across trust boundaries, and what this implies for future defenses.

\section{Blockchain Architecture and Threat Model}
\label{sec:architecture}

To decompose complex Web3 vulnerabilities, this section first establishes a layered architecture model and then defines the adversarial capabilities that recur throughout the survey.

\subsection{Layered Architecture Overview}

Modern blockchains are distributed state machines rather than simple append-only logs \cite{web3sec012_blockchainsvsdistributeddatabasesdichotomy}. Permissioned systems and benchmarking work, including Hyperledger Fabric and BLOCKBENCH, reinforce this systems view \cite{web3sec010_hyperledgerfabricdistributedoperatingsystem,web3sec011_blockbenchframeworkanalyzingprivateblockchains}. Building on empirical vulnerability studies \cite{web3sec014_blockchainsmartcontractsecuritythreats,web3sec015_empiricalstudyblockchainsystemvulnerabilities}, we use four layers:

\begin{itemize}
  \item \textbf{Network layer.} This bottom layer consists of globally distributed nodes connected through P2P protocols. It handles node discovery, connection maintenance, and the propagation of transactions and blocks. It is the physical basis for connectivity and censorship resistance.
  \item \textbf{Data and cryptographic layer.} This layer defines ledger state and the cryptographic primitives that make public verification possible, including authenticated data structures, signatures, and zero-knowledge proofs.
  \item \textbf{Consensus and incentive layer.} Consensus protocols such as proof of work (PoW) and proof of stake (PoS) determine how distributed agreement is reached under Byzantine conditions. Incentive mechanisms, including block rewards and transaction fees, use game-theoretic design to motivate protocol participation.
  \item \textbf{Application layer.} This layer includes Turing-complete smart contracts running on virtual machines such as the EVM, as well as DeFi protocols and DApp frontends composed from such contracts. It is the primary locus of current Web3 value flow and of many logic-level vulnerabilities.
\end{itemize}

Foundational work on the Bitcoin backbone, asynchronous ledgers, PoS, BFT, optimistic confirmation, and sharding formalized the core substrate on which later security analyses build \cite{web3sec002_bitcoinbackboneprotocolanalysisapplications,web3sec003_analysisblockchainprotocolasynchronousnetworks,web3sec004_algorandscalingbyzantineagreementscryptocurrencies,web3sec009_hotstuffbftconsensuslinearityresponsiveness,web3sec008_thunderellablockchainsoptimisticinstantconfirmation,web3sec005_chainspaceshardedsmartcontractsplatform,web3sec006_omniledgersecurescaleoutdecentralized,web3sec007_rapidchainscalingblockchainfullsharding}. We also use \emph{cross-domain boundary} for any point where a protocol imports an external claim and triggers state changes under another trust model, as in oracles, bridges, and rollups.

\subsection{Security Properties as Cross-Layer Invariants}

Layered descriptions are useful, but blockchain failures rarely respect layer boundaries. For that reason, we use five security properties as cross-layer invariants throughout the survey.

\heading{Safety.} Safety means that honest participants do not accept mutually inconsistent histories or invalid state transitions. In a simple payment ledger, safety is violated by double spending. In a smart-contract platform, the same idea extends to invalid execution and forged cross-domain state. Safety is usually stated at the consensus layer, but it depends on lower-layer message delivery and upper-layer validation rules.

\heading{Liveness.} Liveness means that valid transactions can eventually be included and finalized. Throughput alone does not capture it: a chain may continue producing blocks while a security-critical actor is effectively censored. This makes liveness a cross-layer property, because failures can arise from network delay, proposer incentives, fee pressure, or sequencer control.

\heading{Finality.} Finality means that accepted state becomes economically or cryptographically irreversible. PoW chains provide probabilistic finality; BFT and many PoS protocols aim for deterministic finality after quorum certificates. Bridges and rollups complicate this notion because they often import the finality of one domain into another domain with different timing, validator sets, and fault assumptions.

\heading{Privacy.} Privacy means that public verification does not reveal more information than intended. In blockchains, privacy must hold across the ledger, the network, and the user's application workflow. A private transfer protocol can still leak information through timing, address reuse, bridge flows, or DApp frontends.

\heading{Accountability.} Accountablity means that misbehavior can be attributed and penalized. It is weaker than prevention, but many optimistic and slashing-based systems rely on it as their practical security backstop. Accountability is fragile when evidence can be censored, delayed, or made too expensive to submit.

These invariants make defense trade-offs explicit. Private relays reduce mempool leakage and may centralize block construction; ZK bridges remove relayers and introduce proof-system risk; rollups improve scalability and depend on challenge windows, data availability, and sequencer behavior. For each defense, the survey records the strengthened invariant and the residual assumption that remains exposed.

\subsection{Threat Model and Attacker Capabilities}

Classical distributed-systems security often assumes Byzantine nodes that attack availability or consistency without regard to cost. In Web3, adversaries are usually rational actors maximizing return. That distinction is central to MEV, flash-loan attacks, and cross-domain arbitrage.

We distinguish three major capability classes:

\begin{itemize}
    \item \textbf{Network-Level Attackers.} Network-level attackers can interfere with underlying infrastructure. They may control large botnet address pools for Sybil deployment, or, in stronger settings, control autonomous systems or Internet service providers capable of BGP hijacking. Their goal is to isolate target nodes, delay block and transaction propagation, or create information asymmetry that enables double spending or consensus forking.
    \item  \textbf{Consensus-Level Attackers.} Consensus-level attackers are internal protocol participants, such as malicious miners or validators. They may control hash power, stake, or proposer rights without necessarily controlling a majority. Rational attackers exploit this partial control by deviating from the default protocol whenever censorship, reordering, or block withholding yields more value than honest participation.
    \item  \textbf{Application-Level Attackers.} Application-level attackers are the most active and financially damaging class in recent years. They often need only a regular account, enough technical knowledge to compose transactions, and temporary capital obtained through flash loans. Their objective is to turn implementation mistakes or economic-model gaps into direct loss of total value locked (TVL).
\end{itemize}

Attackers may strike within one layer or combine capabilities, such as pairing network delay with application-level arbitrage. The following sections analyze attacks and defenses from the bottom of the stack upward.

The most realistic adversaries are adaptive. They observe liquidity, validator schedules, gas prices, bridge queues, oracle update cadence, and governance timelocks before choosing the cheapest attack path. A protocol that is safe under static assumptions may become unsafe when these variables line up. For example, an attacker may wait for low liquidity, submit transactions through a private relay, target an oracle update boundary, and use a flash loan to make the exploit atomic. We model attacker capability as conditional on market state and timing, in addition to fixed resource thresholds.

\subsection{Adversarial Cost, Timing, and Observability}

The same vulnerability has different implications depending on capital cost, timing, and observability. A consensus bug requiring sustained majority hash power differs from a flash-loan exploit completed in one transaction; a privacy leak found after months of analysis differs from a mempool leak exploitable before finality. These dimensions explain why purely technical defenses are often incomplete: a detector that fires after settlement, a fraud-proof path that challengers cannot afford during congestion, or a privacy scheme that leaks timing at the network layer may preserve the wrong property at the wrong time. Effective defenses must align detection, response, and settlement.

\section{Network Layer Security}
\label{sec:network}

The network layer carries consensus state and application interactions. Public blockchains typically use permissionless P2P overlay networks and gossip protocols to disseminate transactions and blocks. This open topology maximizes decentralization, but it lacks global admission control and authoritative routing validation. As a result, it is structurally vulnerable to partitions, deanonymization, and asymmetric resource exhaustion.

\subsection{Topology and Connection Attacks}

Blockchain security relies heavily on the physical assumption that an honest majority can communicate promptly and without obstruction. Once an adversary can infer or reconstruct the connection graph of underlying nodes \cite{web3sec024_txprobediscoveringbitcoinsnetworktopology}, it can inject malicious routing information, manipulate a victim's network view, and isolate selected nodes.

\heading{Logical isolation: eclipse attacks and topology manipulation.}
Eclipse attacks disconnect a target from all honest peers and place it inside an information environment controlled by the attacker. Heilman et al. showed that Bitcoin nodes could be eclipsed by manipulating the \texttt{Tried/New} address tables so that, after restart or reconnection, outbound connections point to malicious IP addresses \cite{web3sec016_eclipseattacksbitcoinspeerpeer}. In isolation, the victim can receive manipulated difficulty information or delayed blocks, enabling double spending or effective dilution of honest mining power.

As network protocols evolved, topology attacks became more subtle. Ethereum's Kademlia-based discovery mechanism is vulnerable to low-cost Sybil pollution and partitioning because node identifiers are not strongly bound to identity resources \cite{web3sec021_partitioningethereumwithouteclipsingit}. More recent attacks emphasize stealth and low resource requirements. The Erebus attack shows that even modern P2P networks with stricter connection limits can be split by exploiting autonomous-system concentration and handshake-level asymmetries \cite{web3sec022_stealthierpartitioningattackagainstbitcoin}. The same concern now extends beyond Bitcoin and Ethereum: recent NDSS work demonstrates eclipse attacks on Monero's P2P network, showing that privacy-oriented ledgers also inherit topology-level fragility \cite{web3sec214_moneroeclipse}.

\heading{Physical isolation: BGP routing hijacking.}
Network-level adversaries with influence over Internet routing can attack more directly. Apostolaki et al. quantified the concentration of Bitcoin traffic at the autonomous-system layer and showed that a small number of major ISPs often carry a large fraction of mining traffic \cite{web3sec017_hijackingbitcoinroutingattackscryptocurrencies}. By announcing more specific BGP prefixes, adversaries can intercept, modify, or drop blockchain traffic at the routing layer.

Such physical routing disruptions require no cryptographic break and can still manipulate block delivery timing \cite{web3sec019_tamperingdeliveryblockstransactionsbitcoin}. In consensus games, artificial partitions split global mining power into weaker fragments and thereby amplify the relative power of selfish miners.

\heading{Traffic deanonymization and routing defenses.}
Open P2P broadcast also provides fertile ground for tracing transaction origins \cite{web3sec027_deanonymisationclientsbitcoinp2pnetwork}. Dandelion and Dandelion++ mitigate the link between IP addresses and transaction sources by separating dissemination into an anonymity-preserving stem phase and a gossip-based fluff phase \cite{web3sec025_dandelionredesigningbitcoinnetworkanonymity,web3sec026_dandelionpluspluslightweightcryptocurrencynetworkingformal}. Recent Ethereum measurements go further by showing that validator identities can be deanonymized through the P2P layer, converting network metadata into consensus-layer privacy risk \cite{web3sec205_ethereumvalidatordeanonymizing}.

Increasing peer degree alone is an expensive defense because it worsens bandwidth consumption. SABRE instead constructs a relay network using trusted execution environments to defend against BGP attacks \cite{web3sec018_sabreprotectingbitcoinagainstrouting}. Erlay reduces transaction relay bandwidth through efficient set reconciliation while preserving high connectivity \cite{web3sec020_erlayefficienttransactionrelaybitcoin}.

\heading{Defense as topology control.}
Blockchain networking defines part of the attack surface. Peer selection, address gossip, relay policy, and churn determine what an attacker can observe or isolate. The relevant evaluation metrics are honest-peer diversity, capture resistance of local views, tail latency, partition resilience, adversarial churn, and the chance that a target loses all honest outbound paths.

\heading{Network assumptions exported to higher layers.}
Upper layers often import network assumptions silently. Payment channels need justice transactions to reach miners before timeout; rollups need fraud proofs and data-availability samples within a challenge window; DeFi liquidations need keepers to observe state and submit transactions under congestion. A short-lived network attack can pay off if it delays the right message class at the right time. Time-critical applications require operationally independent submission paths and evidence that connectivity survives targeted delay.

Measurement is especially important because average propagation latency hides the tail events that matter for security. A liquidation bot, fraud prover, or channel watchtower fails when its transaction misses a narrow window. Useful measurements include peer diversity, autonomous-system concentration, relay dependence, transaction rebroadcast behavior, and the probability of losing all honest outbound paths. These metrics turn network health into an explicit input for higher-layer risk management.

\subsection{Denial of Service and Bandwidth Exhaustion}

Traditional distributed denial-of-service attacks flood targets with large volumes of meaningless traffic. In blockchains, however, global state transitions are serialized and throughput is constrained by the weakest validating nodes \cite{web3sec038_stabilityscalabilityblockchainsystems,web3sec039_bidlhighthroughputlowlatency,web3sec040_blockenehighthroughputblockchainover}. As a result, crude traffic flooding has evolved into protocol-level asymmetric resource exhaustion.

\heading{Blockchain-specific DoS.}
BDoS attacks combine denial of service with miners' rational behavior \cite{web3sec029_bdosblockchaindenialserviceattacks}. Instead of sending spam, the adversary broadcasts carefully constructed blocks that are extremely expensive to validate. Honest nodes face high sunk validation costs, and rational miners may abandon validation and mine empty blocks to avoid wasting computation. The attack exploits asymmetric validation cost and threatens consensus liveness. Public RPC services also form a fragile point in the Web3 interaction path, since many DApps rely on centralized RPC providers that may lack fine-grained incentives and rate-limiting models \cite{web3sec028_strongitsweakestlinkhow}.

\heading{Resource metering and MPT exploits.}
To avoid nontermination in Turing-complete environments, platforms such as Ethereum and EOSIO introduce gas as a proxy for execution cost \cite{web3sec032_understandingmisbehavioreosioblockchain}. Static gas pricing, however, often drifts away from real validator work. Broken Metre demonstrates that specific EVM bytecode patterns can consume far more CPU time than their gas price suggests, allowing attackers to overload validators at low cost \cite{web3sec031_brokenmetreattackingresourcemetering}. Auspex generalizes this concern to transaction-fee mechanisms, showing that inconsistencies between fee rules and concrete execution behavior can become exploitable consensus and application-level bugs \cite{web3sec201_auspextransactionfee}.

Storage-level exhaustion is even more damaging. The NURGLE attack generates specially prefixed fragmented accounts inside smart contracts to unbalance the Merkle Patricia Trie and force expensive random disk I/O during state lookup \cite{web3sec030_nurgleexacerbatingresourceconsumptionblockchain}. Because each operation complies with gas rules, the attack causes persistent state bloat and undermines gas-limit-based throttling.

\heading{Defensive limits of resource pricing.}
Gas pricing is a coarse scheduling policy: the VM charges abstract operations while validators pay concrete computation, storage, and network costs. When the mapping drifts, attackers search for underpriced execution paths. Durable defenses require measurement feedback, resource profiles for execution traces, and gas schedules treated as empirical parameters rather than constants. State management has the same security role, because old state turns short-term fees into long-term validation debt. The RPC layer creates a parallel asymmetry: applications that rely on a few gateways can be denied service without any consensus attack.

\subsection{Summary and Insights}

Three findings are stable across the network literature. First, topology and routing attacks are mature enough to be modeled and measured, but deployment evidence for peer-diversity policies remains thin. Second, resource-exhaustion attacks exploit mismatches between abstract pricing and concrete validator cost. Third, higher-layer protocols reuse network liveness assumptions without measuring them. The main open measurement gap is tail behavior: how often a target loses all honest paths, misses a challenge window, or depends on a single relay or RPC provider.

\section{Data and Cryptographic Layer Security}
\label{sec:crypto}

The data and cryptographic layer defines ledger state and carries two core trust foundations: unforgeability and privacy. As on-chain analytics mature and zero-knowledge systems become widely deployed, this layer faces both implementation vulnerabilities and increasingly severe privacy leakage.

\subsection{Cryptographic Primitive Vulnerabilities}

Blockchains rely heavily on asymmetric signatures, hash functions, and related primitives. These primitives may be theoretically secure, but flawed implementation or contextual misuse can be disastrous.

\heading{Signature malleability and entropy failures.}
Early Bitcoin exposed transaction malleability: attackers could alter a transaction identifier without invalidating the ECDSA signature. This weakness broke protocols that depended on unconfirmed transaction hashes and contributed to systemic accounting failures in early exchange infrastructure. SegWit mitigated the problem architecturally by separating signature data, but signature-level implementation hazards remain.

ECDSA-style mechanisms require high-entropy nonces. If the same nonce $k$ is reused across transactions, or if the pseudorandom generator is predictable, the private key can be extracted through elementary linear algebra. In decentralized environments, generating unbiased and unpredictable randomness is difficult. Threshold signatures and verifiable random functions have therefore been used to build distributed random beacons for smart-contract execution and committee selection \cite{web3sec068_asynchronousdistributedkeygenerationcomputationally,web3sec069_fullydistributedverifiablerandomfunctions}.

\heading{Under-constrained zero-knowledge circuits.}
Layer-2 scaling and privacy requirements have driven the large-scale deployment of zero-knowledge proofs (ZKPs). Translating high-level logic into low-level polynomial constraint systems, such as R1CS arithmetic circuits or PLONK-style gates, can introduce missing constraints. Under-constrained circuits allow adversaries to provide illegal witnesses while still satisfying the verifier. This breaks soundness and may allow forged proofs, invalid state transitions, or unauthorized asset creation.

The usual failure lies between the intended computation and the algebra actually enforced. Missing constraints over ranges, flags, decompositions, public inputs, or lookups are local engineering mistakes, but rollups and zkVMs can turn them into invalid state transitions. The SoK literature on SNARK vulnerabilities and automated vetting of Polygon zkEVM show that proof-system engineering must be audited with the same adversarial intensity as consensus code and smart contracts \cite{web3sec231_soksnarkvulnerabilities,web3sec210_polygonzkevmfreever}.

Defenses are becoming more circuit-aware. Automated detectors can expose witness variables that vary without changing public outputs \cite{web3sec230_automatedunderconstrainedcircuits}, while refinement types and formal verification can prove that the constraint system implements the intended relation \cite{web3sec064_certifyingzeroknowledgecircuitsrefinement,web3sec232_formalverificationzkcircuits}. High-value rollups also need differential testing across the native implementation, the circuit, and the on-chain verifier. The security target is \emph{end-to-end soundness}: every component agrees on the same state-transition relation.

\heading{The implementation gap in deployed cryptography.}
The most dangerous cryptographic failures often occur between a proven primitive and its deployment. A wallet may reuse nonces; a circuit may omit a range check; a bridge may verify a hash but parse the message differently across chains. Blockchain cryptography is embedded in VMs, DSL compilers, smart contracts, governance, and gas markets, so audits must cover exact encodings, exceptional cases, constraint systems, verifier upgrades, and exit paths. A proof system can be sound yet operationally fragile if proving centralizes, verification becomes unaffordable, or proof data strains availability.

\subsection{Privacy Leakage and Deanonymization}

Decentralization requires transactions to be publicly broadcast and globally verifiable. Early protocol designers often treated pseudonymous public-key-derived addresses as strong anonymity. Quantitative research has refuted this assumption.

\heading{Heuristic graph clustering and tracing.}
Meiklejohn et al. used multi-input co-spend heuristics and change-address inference to cluster Bitcoin addresses and map them to real-world entities such as exchanges and markets \cite{web3sec043_fistfulbitcoinscharacterizingpaymentsamong}. Subsequent work introduced traffic-timing analysis, cross-ledger tracing, and graph-embedding techniques \cite{web3sec044_evaluatinguserprivacybitcoin,web3sec045_analysisanonymitybitcoinusingp2p,web3sec048_tracingtransactionsacrosscryptocurrencyledgers,web3sec049_howpeelmillionvalidatingexpanding}. With platforms such as BlockSci, on-chain deanonymization has moved from theory to industrial-scale practice \cite{web3sec050_blockscidesignapplicationsblockchainanalysis}.

\heading{Fragility of privacy-enhancing protocols.}
Even cryptocurrencies built on advanced privacy primitives remain fragile in practice. Analyses of Zcash and Monero show that nonrandom user behavior, such as repeated transfers between shielded and transparent pools, can sharply reduce anonymity sets \cite{web3sec046_empiricalanalysisanonymityzcash,web3sec047_empiricalanalysislinkabilitymoneroblockchain}. Ground-truth studies of mixing services \cite{web3sec073_mixedsignalsanalyzinggroundtruth} and remote side-channel attacks based on node-response timing \cite{web3sec056_remotesidechannelattacksanonymous} further demonstrate that application-layer mixing alone cannot withstand adversaries with network-wide observation capabilities.

Privacy becomes even harder once assets move across protocols. A transfer that is hidden on one ledger may become linkable through bridge timing, routing behavior, or the intent it exposes to a trading venue. Privacy-preserving multi-hop locks and private atomic-swap protocols reduce some of these cross-protocol linkages \cite{web3sec074_privacypreservingmultihoplocks,web3sec075_sweepucswappingcoinsprivately}, but they do not remove metadata created by liquidity and routing constraints. In practice, privacy is a workflow property: wallet defaults and relayer behavior can matter as much as the cryptographic primitive.

This distinction separates \emph{cryptographic anonymity}, measured on an ideal ledger transcript, from \emph{operational anonymity}, measured under real timing, network location, exchange deposits, bridge exits, and user habits. Many systems satisfy the former and fail the latter.

\subsection{Evolution of Advanced Privacy-Preserving Technologies}

The field has pursued a decade-long cryptographic and systems evolution to reconcile global verifiability with local confidentiality. Table~\ref{tab:privacy-tech} summarizes major privacy-preserving paradigms.

\begin{table*}[t]
\centering
\caption{Comparison of mainstream privacy-preserving technologies.}
\label{tab:privacy-tech}
\scriptsize
\setlength{\tabcolsep}{4.6pt}
\renewcommand{\arraystretch}{1.05}
\setlength{\dashlinedash}{0.65pt}
\setlength{\dashlinegap}{0.95pt}
\setlength{\arrayrulewidth}{0.35pt}
\resizebox{\textwidth}{!}{%
\begin{tabular}{|c|c|c|c|c|c|}
\hline
\textbf{Paradigm} &
\textbf{Representative protocol} &
\textbf{Core cryptography} &
\textbf{Setup} &
\textbf{Cost profile} &
\textbf{Anonymity set} \\
\hline\hline
Network-layer anonymity &
Dandelion \cite{web3sec025_dandelionredesigningbitcoinnetworkanonymity}, Dandelion++ \cite{web3sec026_dandelionpluspluslightweightcryptocurrencynetworkingformal} &
Randomized relay, stem-fluff diffusion &
\nocov &
Low overhead; added propagation latency &
Topology-dependent \\
\hdashline
Decentralized mixers &
CoinJoin \cite{web3sec073_mixedsignalsanalyzinggroundtruth}, Tornado Cash \cite{web3sec073_mixedsignalsanalyzinggroundtruth} &
Hashes, Merkle trees &
\nocov &
Low proof and verification cost &
Pool-limited \\
\hdashline
Ring signatures &
Monero/CryptoNote \cite{web3sec047_empiricalanalysislinkabilitymoneroblockchain}, OmniRing \cite{web3sec057_omniringscalingupprivatepayments} &
Ring signatures, stealth addresses &
\nocov &
Moderate; linear in ring size &
Moderate \\
\hdashline
Anonymous payment channels &
TumbleBit \cite{web3sec053_tumblebituntrustedbitcoincompatibleanonymous}, Bolt \cite{web3sec054_boltanonymouspaymentchannelsdecentralized} &
Blind signatures, HTLCs, ZK proofs &
\nocov &
Interactive; requires collateral and online availability &
Channel- or hub-limited \\
\hdashline
Early ZK-SNARKs &
Zerocash \cite{web3sec042_zerocashdecentralizedanonymouspaymentsbitcoin} &
Pairing-based cryptography &
\fullcov &
Small proofs, fast verification &
Global \\
\hdashline
Transparent ZKPs &
Bulletproofs \cite{web3sec060_bulletproofsshortproofsconfidentialtransactions} &
Inner-product arguments &
\nocov &
Small proofs; slower than SNARKs &
Flexible \\
\hdashline
Confidential transactions &
Mimblewimble/Aggregate Cash \cite{web3sec061_aggregatecashsystemscryptographicinvestigation} &
Pedersen commitments, range proofs &
\nocov &
Moderate; aggregation improves amortized cost &
Transaction-graph reduction \\
\hdashline
Private smart contracts &
Hawk \cite{web3sec052_hawkblockchainmodelcryptographyprivacy} &
ZK proofs, off-chain execution manager &
\partcov &
High design and proving cost &
Application-specific \\
\hdashline
Post-quantum ZKPs &
MatRiCT \cite{web3sec058_matrictefficientscalablepostquantum}, MatRiCT+ \cite{web3sec059_matrictplusmoreefficientpostquantum} &
Lattice-based cryptography &
\nocov &
Larger proofs; moderate verification &
Flexible \\
\hdashline
Hardware-assisted privacy &
BITE \cite{web3sec055_bitebitcoinlightweightclientprivacy} &
Trusted execution environments &
\nocov &
Hardware trust and attestation &
Global \\
\hline
\end{tabular}%
}
\vspace{2pt}
\begin{flushleft}
\scriptsize
Setup coding: \fullcov = required, \partcov = mixed requirement, \nocov = not required. The cost profile summarizes proof size, verification effort, and deployment assumptions rather than asymptotic complexity alone.
\end{flushleft}
\end{table*}

\heading{From mixers to noninteractive zero knowledge.}
Early privacy systems used multisignatures and HTLCs for trust-minimized payment hubs, as in TumbleBit and Bolt \cite{web3sec053_tumblebituntrustedbitcoincompatibleanonymous,web3sec054_boltanonymouspaymentchannelsdecentralized}, or ring signatures to hide spenders, as in Omniring \cite{web3sec057_omniringscalingupprivatepayments}. Zerocoin and Zerocash made noninteractive anonymous transfer possible with zk-SNARKs \cite{web3sec041_zerocoinanonymousdistributedecash,web3sec042_zerocashdecentralizedanonymouspaymentsbitcoin}, while Bulletproofs, Mimblewimble-style aggregate cash, improved range arguments, and vector-commitment aggregation reduced setup or proof-size costs \cite{web3sec060_bulletproofsshortproofsconfidentialtransactions,web3sec061_aggregatecashsystemscryptographicinvestigation,web3sec062_efficientzeroknowledgeargumentspaillier,web3sec063_swiftrangeshortefficientzeroknowledge,web3sec067_pointproofsaggregatingproofsmultiplevector}. Hawk extended privacy to programmable contracts, and TEE-based light clients improved privacy for constrained users \cite{web3sec052_hawkblockchainmodelcryptographyprivacy,web3sec055_bitebitcoinlightweightclientprivacy}.

\heading{Post-quantum readiness and parallel scaling.}
Three trends now dominate: lattice-based confidential transactions for post-quantum readiness \cite{web3sec058_matrictefficientscalablepostquantum,web3sec059_matrictplusmoreefficientpostquantum}, distributed ZK proving to reduce prover bottlenecks \cite{web3sec066_pianistscalablezkrollupsfullydistributed,web3sec065_ligetronlightweightscalableendend}, and certified circuits or privacy-aware PoS designs that reduce implementation and metadata leakage \cite{web3sec064_certifyingzeroknowledgecircuitsrefinement,web3sec070_ouroboroscrypsinousprivacypreservingproof,web3sec071_proofstakeprotocolsprivacyaware,web3sec072_anonymityguaranteesanonymousproofstake}.

\subsection{Summary and Insights}

The cryptographic literature is strong on primitive design and weaker on deployed workflows. ZK-SNARKs, confidential transactions, and post-quantum privacy schemes address confidentiality, yet many failures arise from nonce reuse, missing circuit constraints, verifier upgrades, or user behavior that shrinks anonymity sets. The unresolved comparison is operational: which privacy systems retain anonymity under exchange flows, bridge timing, relayer metadata, and compliance checks? Proofs of solvency and lawful-provenance proofs provide one concrete test case for this question \cite{web3sec051_provisionsprivacypreservingproofssolvency}.

\section{Consensus and Incentive Layer Security}
\label{sec:consensus}

Classical fault-tolerant distributed systems mainly analyze crash or Byzantine behavior. In tokenized Web3 systems, nodes with protocol weight have strong economic incentives to exploit edge cases. The threat model shifts from pure resource domination to timing manipulation and strategic games. Table~\ref{tab:consensus-threats} summarizes representative threats.

\begin{table*}[t]
\centering
\caption{Consensus-layer attacks and their interaction with execution-layer value.}
\label{tab:consensus-threats}
\scriptsize
\setlength{\tabcolsep}{3.7pt}
\renewcommand{\arraystretch}{1.06}
\setlength{\dashlinedash}{0.65pt}
\setlength{\dashlinegap}{0.95pt}
\setlength{\arrayrulewidth}{0.35pt}
\resizebox{\textwidth}{!}{%
\begin{tabular}{|c|cc|cc|c|c|}
\hline
\multirow{2}{*}{\textbf{Attack vector}} &
\multicolumn{2}{c|}{\textbf{Targeted blockchain}} &
\multicolumn{2}{c|}{\textbf{Attack surface}} &
\multirow{2}{*}{\textbf{Attack impact}} &
\multirow{2}{*}{\textbf{Representative defense}} \\
\cline{2-5}
&
\textbf{Consensus} &
\textbf{Implementation} &
\textbf{Consensus layer} &
\textbf{Execution layer} &
&
\\
\hline\hline
Selfish mining \cite{web3sec076_majoritynotenoughbitcoinmining} &
PoW &
Mining pools &
\fullcov &
\nocov &
Disproportionate rewards &
Uniform tie-breaking; Fruitchains \cite{web3sec082_fruitchainsfairblockchain} \\
\hdashline
Block withholding / pool infiltration \cite{web3sec077_minersdilemma} &
PoW &
Mining pools &
\fullcov &
\nocov &
Pool revenue loss; miner's dilemma &
Decentralized pools; reward auditing \cite{web3sec080_smartpoolpracticaldecentralizedpooledmining} \\
\hdashline
Bribery mining \cite{web3sec222_novelbriberymining} &
PoW &
External bribery markets &
\fullcov &
\nocov &
Coalition formation; incentive distortion &
Bribery-resistant reward design; monitoring \\
\hdashline
Mining reward exhaustion / halt game \cite{web3sec218_haltgamepow} &
PoW &
Fee-and-reward market &
\fullcov &
\nocov &
Miner exit; liveness degradation &
Sustainable fee markets; adaptive rewards \\
\hdashline
Fee sniping / fee instability \cite{web3sec079_instabilitybitcoinwithoutblockreward} &
PoW &
High-fee blocks &
\fullcov &
\fullcov &
Forking to capture transaction fees &
Stable subsidy design; fee smoothing \\
\hdashline
Timestamp manipulation \cite{web3sec220_timestampmanipulation} &
Nakamoto-style &
Timestamp-based rules &
\fullcov &
\fullcov &
Biased ordering; invalidated competitors &
Tighter timestamp validation; delay-aware rules \\
\hdashline
Uncle Maker \cite{web3sec119_unclemakertimestampingout} &
Ethereum PoW &
Uncle-block mechanism &
\fullcov &
\fullcov &
Competitor block invalidation &
Timestamp hardening; fork-choice revision \\
\hdashline
Long-range attack \cite{web3sec242_winklefoilongrange} &
PoS &
Validator keys &
\fullcov &
\nocov &
Ledger history overwrite &
Key-evolving signatures; checkpoints \cite{web3sec085_ouroborospraosadaptivelysecuresemi} \\
\hdashline
Nothing-at-stake / equivocation \cite{web3sec243_formalbarrierslongestchainpos} &
PoS &
Validator voting &
\fullcov &
\nocov &
Conflicting histories; finality delay &
Slashing; accountable safety \cite{web3sec084_ouroborosprovablysecureproofstake} \\
\hdashline
RANDAO grinding / randomness manipulation \cite{web3sec239_optimalrandaomanipulationethereum,web3sec240_forkingrandao} &
Ethereum PoS &
Randomness beacon &
\fullcov &
\fullcov &
Biased proposer schedules; future ordering control &
VDFs; threshold randomness; delayed entropy \\
\hdashline
Balancing / bouncing attacks \cite{web3sec093_ebbflowprotocolsresolutionavailability,web3sec235_threeattacksposethereum,web3sec236_twomoreattackspostghostethereum,web3sec238_ethereumproofstakeunderscrutiny} &
Ethereum PoS &
LMD-GHOST/FFG timing &
\fullcov &
\nocov &
Finality delay; liveness loss &
Safe slots; Goldfish \cite{web3sec241_goldfishnomoreattacksethereum} \\
\hdashline
Ex-ante / sandwich reorganization \cite{web3sec237_lowcostattacksethereum20,web3sec235_threeattacksposethereum,web3sec202_availableattestation} &
Ethereum PoS &
Proposer boost and block weights &
\fullcov &
\fullcov &
Honest block orphaning; MEV-driven reorgs &
Available attestation; Goldfish \cite{web3sec202_availableattestation,web3sec241_goldfishnomoreattacksethereum} \\
\hdashline
Justification withholding / unrealized justification \cite{web3sec235_threeattacksposethereum,web3sec236_twomoreattackspostghostethereum,web3sec202_availableattestation} &
Ethereum PoS &
Checkpoint filtering &
\fullcov &
\nocov &
Canonical-chain pruning; finality stalls &
Capella/Deneb hardening; available attestation \cite{web3sec202_availableattestation} \\
\hdashline
Staircase attestation-incentive attack \cite{web3sec234_maxattestationmatters} &
Ethereum PoS &
Attestation inclusion limits &
\fullcov &
\nocov &
Honest-validator penalties; stake drift &
Parameter tuning; upgrade hardening \\
\hdashline
No-cost liveness / strong-liveness attack \cite{web3sec233_livenessattackethereumposnoadditionalcost} &
Ethereum PoS &
RANDAO and checkpoint filtering &
\fullcov &
\fullcov &
No finality; Byzantine-only finalized blocks &
RANDAO lookahead limits; checkpoint redesign \\
\hdashline
Finality-gadget failure \cite{web3sec204_bscfinalitygadget} &
PoS/BFT hybrid &
Finality gadget &
\fullcov &
\nocov &
False finality; rollback risk &
Formal finality checks; safer fork choice \\
\hdashline
BGP routing hijack \cite{web3sec017_hijackingbitcoinroutingattackscryptocurrencies} &
PoW/PoS &
AS-level routing &
\fullcov &
\nocov &
Delayed block propagation &
Relay networks such as SABRE \cite{web3sec018_sabreprotectingbitcoinagainstrouting} \\
\hdashline
Eclipse / partition attack \cite{web3sec016_eclipseattacksbitcoinspeerpeer,web3sec021_partitioningethereumwithouteclipsingit} &
PoW/PoS &
P2P topology &
\fullcov &
\nocov &
Victim isolation; double-spend support &
Peer diversity; hardened discovery \\
\hdashline
Blockchain DoS / validation exhaustion \cite{web3sec029_bdosblockchaindenialserviceattacks} &
PoW/PoS &
Block validation path &
\fullcov &
\fullcov &
Consensus liveness loss &
Resource pricing; validation-cost limits \\
\hdashline
Time-bandit attack \cite{web3sec111_flashboys2frontrunningdecentralized} &
Any &
MEV-rich execution market &
\fullcov &
\fullcov &
Reorganization to steal past MEV &
Fair sequencing; PBS; MEV smoothing \\
\hdashline
Front-running / ordering manipulation \cite{web3sec177_sokpreventingtransactionreorderingmanipulations} &
Any &
Mempool and proposer ordering &
\fullcov &
\fullcov &
Sandwiching; liquidation theft &
Commit-reveal; batch auctions; encrypted mempools \\
\hdashline
Cross-shard front-running \cite{web3sec110_frontrunningattackshardedblockchains} &
Sharded chains &
Cross-shard ordering &
\fullcov &
\fullcov &
Cross-domain ordering advantage &
Fair cross-shard consensus \\
\hline
\end{tabular}%
}
\vspace{2pt}
\begin{flushleft}
\scriptsize
\fullcov = required; \nocov = not required.
\textbf{Consensus layer} denotes dependence on consensus state, voting, fork choice, or stability/finality predicates.
\textbf{Execution layer} denotes dependence on payload semantics, transaction ordering, application-layer value, or proposer/builder capture.
\end{flushleft}
\end{table*}

\subsection{Consensus Rule Attacks and Theoretical Bounds}

The most direct attack is to aggregate enough physical or economic resources to reverse finality or block liveness.

\heading{Security bounds and scaling of PoW.}
Nakamoto consensus relies on honest majority. Formal analyses characterize safety bounds and confirmation depth under network delay \cite{web3sec003_analysisblockchainprotocolasynchronousnetworks,web3sec078_securityperformanceproofworkblockchains,web3sec088_laydowncommonmetricsevaluating,web3sec090_everythingracenakamotoalwayswins,web3sec091_tightconsistencyboundsbitcoin,web3sec092_revisitingnakamotoconsensusasynchronousnetworks}. Fee-only systems can become unstable because miners may fork to capture high fees \cite{web3sec079_instabilitybitcoinwithoutblockreward}; the Halt Game further shows that when rewards fail to cover expenses, rational miners may stop participating, turning economic sustainability into a liveness condition \cite{web3sec218_haltgamepow}. Scale has similarly nuanced effects: larger Nakamoto-style systems may or may not improve resilience depending on delay and resource-distribution assumptions \cite{web3sec120_largerscalenakamotostyleblockchains,web3sec224_largerscaleoffersbettersecurity}.

Performance-oriented designs separate leader election from serialization, reduce block-production variance, or diversify scarce consensus resources \cite{web3sec097_bitcoinngscalableblockchainprotocol,web3sec098_prismdeconstructingblockchainapproachphysical,web3sec094_ncmaxbreakingsecurityperformance,web3sec095_bobtailimprovedblockchainsecuritylow,web3sec096_strongchaintransparentcollaborativeproofwork,web3sec089_ohieblockchainscalingmadesimple,web3sec099_minotaurmultiresourceblockchainconsensus}. They improve throughput but shift incentives: Bitcoin-NG, for example, raises throughput by separating key blocks from microblocks, while Greedy-Mine shows that the split can be manipulated for profit \cite{web3sec221_greedyminebitcoinng}.

\heading{Endogenous flaws of PoS and asynchronous BFT.}
PoS replaces physical energy with on-chain weight and introduces attacks such as nothing-at-stake behavior and long-range history fabrication. The Ouroboros family, Snow White, and PoS sidechain work provide formal security under carefully stated synchrony, leader-election, and key-evolution assumptions \cite{web3sec084_ouroborosprovablysecureproofstake,web3sec085_ouroborospraosadaptivelysecuresemi,web3sec086_ouroborosgenesiscomposableproofstake,web3sec083_snowwhiterobustlyreconfigurableconsensus,web3sec087_proofstakesidechains}. Ethereum PoS research shows that attestation availability, fork-choice timing, validator incentives, and RANDAO entropy remain active surfaces: low-cost reorganization, balancing, bouncing, justification-withholding, unrealized-justification, staircase, liveness, and randomness-manipulation attacks can degrade liveness, chain quality, or incentives without obvious slashing behavior \cite{web3sec237_lowcostattacksethereum20,web3sec235_threeattacksposethereum,web3sec236_twomoreattackspostghostethereum,web3sec238_ethereumproofstakeunderscrutiny,web3sec234_maxattestationmatters,web3sec233_livenessattackethereumposnoadditionalcost,web3sec239_optimalrandaomanipulationethereum,web3sec240_forkingrandao}. These results motivate reorg-resilient attestations, Goldfish-style fork choice, and automated search for incentive flaws \cite{web3sec202_availableattestation,web3sec241_goldfishnomoreattacksethereum,web3sec203_bunnyfinderethereumconsensus}.

\heading{Restaking and shared-security spillovers.}
Restaking lets the same collateral secure multiple actively validated services (AVSs). This improves capital efficiency but turns validator risk into a shared balance sheet. Robust-restaking and liquid-restaking analyses show that stake allocation, slashing rules, operator overlap, derivative leverage, and adversarial payoff jointly determine whether one AVS failure spills into another \cite{web3sec225_robustrestakingnetworks,web3sec226_financialdynamicsrestaking}. The risk is \emph{consensus overload}: shared collateral needs explicit loss containment, and additional reuse can increase correlated loss.

High-concurrency BFT protocols increasingly decouple data dissemination from metadata ordering. DAG-based and asynchronous designs improve throughput while preserving classical safety and liveness goals \cite{web3sec100_bullsharkdagbftprotocolsmade,web3sec101_narwhaltuskdagbasedmempool,web3sec102_redbellysecurefairscalable,web3sec103_dumbofasterasynchronousbftprotocols,web3sec104_dumbongfastasynchronousbft,web3sec105_synchotstuffsimplepracticalsynchronous}; ebb-and-flow protocols show that availability and finality must be designed together \cite{web3sec093_ebbflowprotocolsresolutionavailability}.

\heading{Finality as an economic interface.}
Applications consume finality as an economic interface. Once a bridge, rollup, or lending market treats state as final, it may release value elsewhere. Probabilistic finality gives a risk curve; deterministic BFT finality gives a sharper interface but depends on quorum and accountability. Hybrid systems need precise rules for disagreement between fast confirmation and slow settlement, a gap highlighted by analysis of Binance Smart Chain's finality gadget \cite{web3sec204_bscfinalitygadget}.

This distinction matters because applications often convert finality into credit. A bridge that releases wrapped assets, a lending market that accepts collateral, or a rollup that honors a withdrawal all turn a consensus judgment into an economic promise. If the application accepts a weaker notion of finality than users believe, the gap becomes a source of reorganization, censorship, or insolvency risk.

\heading{Committee selection and adaptive corruption.}
PoS and BFT committees reduce communication cost but turn security into a sampling problem. VRFs and distributed randomness reduce manipulation, yet early membership revelation or targetable validator keys still enable adaptive corruption. Committee security spans selection secrecy, key management, and post-failure accountability.

\subsection{Incentive Misalignment and Strategic Mining}

Even below the 51\% threshold, rational nodes may exploit reward-allocation flaws.

\heading{Game-theoretic analysis of selfish mining.}
Eyal and Sirer showed that honest mining is not always dominant: a pool with enough hash power and propagation advantage can withhold blocks and release them strategically to earn more than its proportional reward \cite{web3sec076_majoritynotenoughbitcoinmining}. This finding led to studies of miner's dilemmas, pool infiltration, and resource-efficient mining \cite{web3sec077_minersdilemma,web3sec080_smartpoolpracticaldecentralizedpooledmining,web3sec081_remresourceefficientminingblockchains}. Recent work extends the line to profitable mining behavior, external bribery, and incentive-compatible coalition formation \cite{web3sec219_bmpawprofitablemining,web3sec222_novelbriberymining,web3sec223_collapsingminersdilemma}. Fruitchains responds by separating block-packaging rewards from transaction-recording rewards, reducing the payoff from withholding \cite{web3sec082_fruitchainsfairblockchain}.

\heading{Perturbations by fee mechanisms.}
Fee mechanisms also affect consensus stability. EIP-1559 changed miners' mempool strategies and short-term security during congestion \cite{web3sec118_empiricalanalysiseip1559transaction}; timestamp tolerance enables Uncle Maker and broader timestamp-manipulation attacks that expose safety and fairness to clock-selection incentives \cite{web3sec119_unclemakertimestampingout,web3sec220_timestampmanipulation}. Fees also determine who can respond under stress: if challengers or liquidators are priced out, the chain remains live while higher-layer safety fails.

\subsection{Maximal Extractable Value: The Dark Forest}

Maximal extractable value (MEV), originally studied as miner extractable value in PoW systems, is the most prominent threat connecting consensus and application layers. Whereas the attacks above target endogenous block rewards, MEV allows miners, validators, builders, or sequencers to use transaction-inclusion and ordering rights to extract application-layer profits.

\begin{figure}[t]
  \centering
  \surveyfig{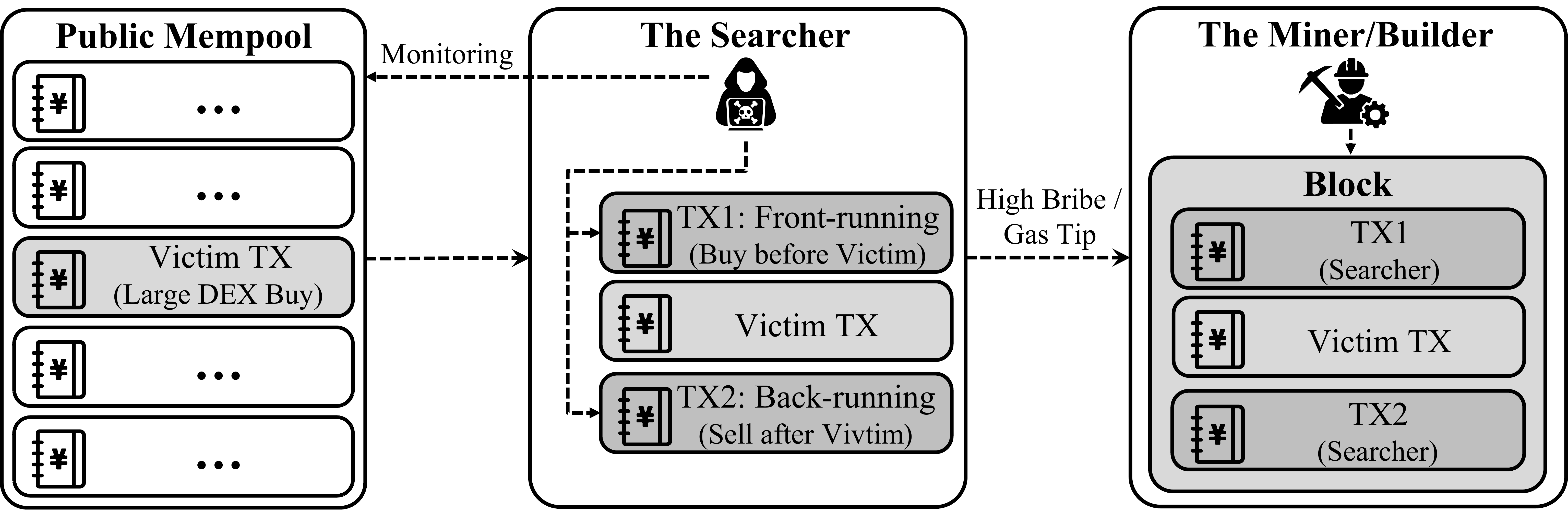}
  \caption{MEV extraction through transaction ordering.}
  \Description{A schematic of MEV extraction in which a searcher observes a victim transaction and submits a bundle that lets a proposer order transactions for sandwich profit.}
  \label{fig:mev}
\end{figure}

\heading{Ruthless extraction and time-bandits.}
\emph{Flash Boys 2.0} systematically quantified high-frequency front-running on decentralized exchanges \cite{web3sec111_flashboys2frontrunningdecentralized}. As Figure~\ref{fig:mev} illustrates, MEV extraction is fundamentally an abuse of ordering power. Searchers monitor public mempools; when they detect a large trade that will move prices, they submit transaction pairs with high priority fees so that miners place them before and after the victim transaction, creating a sandwich attack.

Large-scale measurement shows that MEV has become a multibillion-dollar shadow economy \cite{web3sec112_highfrequencytradingdecentralizedchain,web3sec113_justtimediscoveryprofitgenerating,web3sec114_quantifyingblockchainextractablevaluehow}. To avoid public gas wars, searchers increasingly submit bundles through private channels such as Flashbots. These dark pools reduce P2P broadcast overhead but centralize block construction and increase censorship risk \cite{web3sec115_flashbotpanmeasuringmaximal,web3sec116_demystifyingdefimevactivitiesflashbots}. MEVisor shows that scalable symbolic reasoning can automate high-throughput MEV discovery \cite{web3sec117_mevisorhighthroughputmevdiscovery}, while recent lifecycle analysis of MEV bots reveals how extraction strategies evolve across discovery, simulation, deployment, and retirement phases \cite{web3sec212_lightdarknessmevbot}.

The time-bandit attack is even more systemic. If the MEV contained in a past block exceeds the protocol block reward, rational miners may be incentivized to reorganize the chain and steal the historical opportunity, undermining immutability.

\heading{Toward fair sequencing.}
MEV points to a gap in classical consensus specifications: they guarantee state consistency and liveness while leaving transaction-order fairness mostly unconstrained. Order-fairness and threshold-encryption protocols limit a proposer's ability to see or dictate the profitable order \cite{web3sec106_themisfaststrongorderfairness,web3sec107_orderfairnessbyzantineconsensus,web3sec108_wendygoodlittlefairnesswidget,web3sec109_separationgoodfasterorderfairness,web3sec110_frontrunningattackshardedblockchains}. Newer systems such as BEAT-MEV and batched threshold encryption indicate that mempool privacy is moving from an abstract fairness goal toward a deployable mitigation primitive \cite{web3sec206_beatmev,web3sec207_mempoolprivacybte}.

Ethereum and related ecosystems increasingly explore proposer-builder separation (PBS), which decouples block construction from proposal rights. PBS can make MEV markets more explicit, while concentration moves toward builders, relays, private order flow, and simulation advantage. Cryptographic ordering tools reduce proposer discretion and add latency and assumptions. MEV mitigation is best evaluated as protocol-level market design.

A complete MEV defense also has to decide where value goes. Some approaches hide transactions, some auction ordering rights, some redistribute extracted value, and some redesign applications to reduce extractable surplus. These choices protect different parties. Encrypted mempools help users and can delay price discovery. PBS can smooth validator revenue and may centralize block construction. Batch auctions reduce sandwiching and require applications to accept coarser execution. The design target can be stated explicitly: reducing user loss, reducing reorg incentives, limiting censorship power, or making extraction auditable.

\subsection{Summary and Insights}

The consensus literature now separates three attack surfaces. Resource-threshold attacks remain foundational; incentive attacks explain deviations below majority control; ordering attacks connect consensus rights to application-layer value. The strongest deployment evidence exists for classic PoW/PoS bounds and MEV measurement, while restaking risk, PBS centralization, and application interpretation of finality still lack long-run empirical evidence. A useful consensus evaluation now reports both safety/liveness predicates and the economic payload that makes deviation profitable.

\section{Application Layer Security}
\label{sec:application}

Application-layer security depends on deployed smart-contract code and the economic invariants it is supposed to enforce. Smart contracts resemble ordinary software, but they execute under an adversarial scheduler and often hold bearer assets that cannot be rolled back. We therefore distinguish code-level vulnerabilities from financial-logic flaws: many damaging transactions are valid calls that exploit the gap between intended invariant and enforced state.

\subsection{Code-Level Vulnerabilities and Automated Analysis}

Since The DAO incident, the security community has systematically classified and analyzed Solidity/EVM execution flaws.

\heading{Landscape of code-level exploits.}
Early taxonomies defined the core vulnerability families that still organize smart-contract auditing \cite{web3sec121_makingsmartcontractssmarter,web3sec122_surveyattacksethereumsmartcontracts,web3sec125_findinggreedyprodigalsuicidalcontracts,web3sec131_artscamdemystifyinghoneypotsethereum,web3sec150_smartcontractvulnerabilitiesvulnerabledoes}. Reentrancy remains the canonical example because an external callback can turn a local state-update order into repeated asset withdrawal \cite{web3sec126_teethergnawingethereumautomaticallyexploit}; newer anti-reentrancy studies show that detectors must distinguish exploitable callbacks from defensive idioms \cite{web3sec216_antireentrancypatterns}. Other work shows the same pattern at different abstraction levels: arithmetic and gas bugs expose low-level execution assumptions \cite{web3sec127_madmaxsurvivingoutgasconditions,web3sec156_etainterdetectinggasrelatedvulnerabilities,web3sec031_brokenmetreattackingresourcemetering,web3sec154_soltyperefinementtypesarithmeticoverflow}, while proxy storage, permission mining, and token-approval analyses expose long-lived authority assumptions \cite{web3sec173_notyourtypedetectingstorage,web3sec158_findingpermissionbugssmartcontracts,web3sec209_approvetransferfrom}. Longitudinal studies show why this remains a moving target: attackers keep disguising known bug patterns inside new protocol contexts \cite{web3sec141_everevolvinggameevaluationreal,web3sec149_evilundersununderstandingdiscovering}.

\heading{From symbolic execution to semantic verification.}
Early automated tools used symbolic execution to pursue exhaustive path coverage. Oyente, ZEUS, and Vandal pioneered abstract code verification \cite{web3sec121_makingsmartcontractssmarter,web3sec123_zeusanalyzingsafetysmartcontracts,web3sec128_vandalscalablesecurityanalysisframework}, while Securify introduced precise Datalog-based data-flow reasoning \cite{web3sec124_securifypracticalsecurityanalysissmart}. Because smart contracts interact heavily with external state, symbolic execution suffers from path explosion.

Later work improved intermediate representations and global execution-context modeling. IR-based analyzers use decompilation, abstraction, and formal reasoning to reduce false positives \cite{web3sec134_gigahorsethoroughdeclarativedecompilationsmart,web3sec135_ethorpracticalprovablysoundstatic,web3sec136_verxsafetyverificationsmartcontracts,web3sec137_verismarthighlyprecisesafetyverifier}, while semantic models and bounded model checking strengthen verification guarantees \cite{web3sec138_semanticunderstandingsmartcontractsexecutable,web3sec139_ethbmcboundedmodelcheckersmart}. Other tools move the analysis boundary outward: TXSPECTOR reasons about real transaction traces \cite{web3sec140_txspectoruncoveringattacksethereumtransactions}, Park and SAILFISH address scale and state inconsistency \cite{web3sec159_parkacceleratingsmartcontractvulnerability,web3sec155_sailfishvettingsmartcontractstate}, and platform-specific analyzers extend auditing beyond the EVM \cite{web3sec157_wasaiuncoveringvulnerabilitieswasmsmart,web3sec160_pandasecurityanalysisalgorandsmart}. For already-deployed contracts, runtime blocking and patching systems attempt to intercept malicious execution or synthesize defenses after deployment \cite{web3sec130_sereumsethprotectingexistingsmart,web3sec142_sodagenericonlinedetectionframework,web3sec146_smartpulseautomatedcheckingtemporalproperties,web3sec147_sguardsmartcontractsmadevulnerability,web3sec148_evmpatchtimelyautomatedpatchingethereum}.

\heading{Rise of fuzzing and AI-driven analysis.}
High false-positive rates in static analysis have moved the field toward dynamic fuzzing. Early fuzzers generated random calls, while later greybox and hybrid systems use path feedback and data dependencies to reach deeper contract states \cite{web3sec129_contractfuzzerfuzzingsmartcontractsvulnerability,web3sec144_sfuzzefficientadaptivefuzzersolidity,web3sec145_harveygreyboxfuzzersmartcontracts,web3sec151_smartianenhancingsmartcontractfuzzing,web3sec152_confuzziusdatadependencyawarehybrid}. Newer work extends this idea with learned guidance and chain-specific execution models \cite{web3sec132_learningfuzzsymbolicexecutionapplication,web3sec172_fuzzbeachfuzzingsolanasmart}. Practical evaluation now considers both coverage and whether a tool produces findings that auditors can act on \cite{web3sec165_largescalestudyvulnerabilityscanners,web3sec167_wethereyetunravelingstate,web3sec168_smartcontractdefisecuritytools}.

Deep-learning-based analysis is now emerging. Graph-neural-network scanners extract structural features from smart contracts \cite{web3sec161_smartercontractsdetectingvulnerabilitiessmart,web3sec166_scvhuntersmartcontractvulnerabilitydetection}, and GPTScan combines LLM-based semantic reconstruction with static analysis \cite{web3sec169_gptscandetectinglogicvulnerabilitiessmart}. Table~\ref{tab:contract-tools} summarizes representative tools.

\begin{table*}[t]
\centering
\caption{Representative smart-contract analysis tools.}
\label{tab:contract-tools}
\scriptsize
\setlength{\tabcolsep}{4.6pt}
\renewcommand{\arraystretch}{1.05}
\setlength{\dashlinedash}{0.65pt}
\setlength{\dashlinegap}{0.95pt}
\setlength{\arrayrulewidth}{0.35pt}
\resizebox{\textwidth}{!}{%
\begin{tabular}{|c|c|c|c|c|}
\hline
\textbf{Tool} &
\textbf{Target platform} &
\textbf{Core technique} &
\textbf{Detected vulnerabilities} &
\textbf{Open source} \\
\hline\hline
Oyente \cite{web3sec121_makingsmartcontractssmarter} &
Ethereum &
Symbolic execution &
Reentrancy, transaction ordering &
\nocov \\
\hdashline
ZEUS \cite{web3sec123_zeusanalyzingsafetysmartcontracts} &
Ethereum &
Abstract interpretation + model checking &
Safety-policy violations &
\fullcov \\
\hdashline
Securify \cite{web3sec124_securifypracticalsecurityanalysissmart} &
Ethereum &
Datalog/static analysis &
DAO-style bugs, unhandled exceptions &
\fullcov \\
\hdashline
MadMax \cite{web3sec127_madmaxsurvivingoutgasconditions} &
Ethereum &
Static analysis &
Gas-related vulnerabilities &
\fullcov \\
\hdashline
teEther \cite{web3sec126_teethergnawingethereumautomaticallyexploit} &
Ethereum &
Exploit generation &
Money-leaking contracts &
\fullcov \\
\hdashline
Sereum \cite{web3sec130_sereumsethprotectingexistingsmart} &
Ethereum &
Runtime monitoring &
Reentrancy &
\fullcov \\
\hdashline
ContractFuzzer \cite{web3sec129_contractfuzzerfuzzingsmartcontractsvulnerability} &
Ethereum &
Black-box fuzzing &
Reentrancy, timestamp, gasless send &
\fullcov \\
\hdashline
sFuzz \cite{web3sec144_sfuzzefficientadaptivefuzzersolidity} &
Ethereum &
Adaptive fuzzing &
Solidity vulnerability patterns &
\fullcov \\
\hdashline
Harvey \cite{web3sec145_harveygreyboxfuzzersmartcontracts} &
Ethereum &
Greybox fuzzing &
Assertions, custom invariants &
\nocov \\
\hdashline
ConFuzzius \cite{web3sec152_confuzziusdatadependencyawarehybrid} &
Ethereum &
Hybrid fuzzing + symbolic execution &
Deep state-dependent bugs &
\fullcov \\
\hdashline
VerX \cite{web3sec136_verxsafetyverificationsmartcontracts} &
Ethereum &
Temporal-property verification &
Contract safety properties &
\nocov \\
\hdashline
VeriSmart \cite{web3sec137_verismarthighlyprecisesafetyverifier} &
Ethereum &
Precise static verification &
Arithmetic and safety bugs &
\fullcov \\
\hdashline
TXSPECTOR \cite{web3sec140_txspectoruncoveringattacksethereumtransactions} &
Ethereum &
Transaction-trace analysis &
Attack patterns in historical traces &
\fullcov \\
\hdashline
SAILFISH \cite{web3sec155_sailfishvettingsmartcontractstate} &
Ethereum &
State inconsistency analysis &
DeFi state-logic flaws &
\fullcov \\
\hdashline
DArcher \cite{web3sec153_darcherdetectingchainoffchain} &
DApps &
On-chain/off-chain differential analysis &
Frontend synchronization bugs &
\fullcov \\
\hdashline
WASAI \cite{web3sec157_wasaiuncoveringvulnerabilitieswasmsmart} &
Wasm contracts &
Static and symbolic analysis &
Wasm smart-contract bugs &
\fullcov \\
\hdashline
Panda \cite{web3sec160_pandasecurityanalysisalgorandsmart} &
Algorand &
Symbolic execution &
TEAL state manipulation &
\fullcov \\
\hdashline
Nyx \cite{web3sec164_nyxdetectingexploitablefrontrunning} &
Ethereum &
Static analysis + exploitability checking &
Front-running vulnerabilities &
\fullcov \\
\hdashline
FlashSyn \cite{web3sec171_flashsynflashloanattacksynthesis} &
Multi-chain DeFi &
Counterexample synthesis &
Flash-loan composability &
\fullcov \\
\hdashline
SCVHunter \cite{web3sec166_scvhuntersmartcontractvulnerabilitydetection} &
Ethereum &
Graph neural network &
Smart-contract vulnerability patterns &
\fullcov \\
\hdashline
GPTScan \cite{web3sec169_gptscandetectinglogicvulnerabilitiessmart} &
Ethereum &
LLM + static analysis &
Financial-logic bugs &
\fullcov \\
\hdashline
HOUSTON \cite{web3sec180_houstonrealtimeanomalydetection} &
Ethereum DeFi &
Real-time anomaly detection &
DeFi attacks in transaction streams &
\nocov \\
\hline
\end{tabular}%
}
\vspace{2pt}
\begin{flushleft}
\scriptsize
The table emphasizes representative techniques rather than exhaustive tool coverage. Open-source coding follows the availability reported by the cited work or public release status: \fullcov = available, \partcov = mixed availability, \nocov = unavailable.
\end{flushleft}
\end{table*}

\heading{Overhead and industrial adoption.}
Despite strong academic results, many tools are not widely adopted in production audits. The obstacle is audit usefulness: practical tools must reduce false alarms, handle external state, explain the required exploit state, estimate feasibility under liquidity and gas constraints, and produce actionable fixes. Empirical evaluations on real contracts, known exploits, and practitioner workflows carry the same weight as new detection algorithms \cite{web3sec165_largescalestudyvulnerabilityscanners,web3sec167_wethereyetunravelingstate,web3sec168_smartcontractdefisecuritytools}. Otherwise automated analysis can report many local issues while missing the economically meaningful attack path.

The key gap is between a warning and an exploit. A reentrancy pattern may be unreachable, an overflow may be guarded by an upstream invariant, and an apparent authorization bug may require an impossible state. Conversely, a syntactically clean contract can be exploitable when composed with a lending market, an oracle, and a flash-loan provider. A useful evaluation rewards tools that explain the path from local condition to feasible loss, rather than tools that only maximize benchmark findings.

\heading{Runtime defenses and the patching dilemma.}
Runtime monitors and automatic patching help because many vulnerable contracts are already deployed, but they create governance risk. A call blocker must define maliciousness under incomplete information, and a patching system must change behavior without becoming a hidden administrator. Such defenses work best for narrow invariants paired with transparent incident response and credible exits.

\subsection{DeFi Protocol and Financial Logic Attacks}

Recent SoK studies and on-chain imitation-game analyses show that many contracts with no apparent code defect under fuzzing still rely on fragile economic assumptions \cite{web3sec174_sokdecentralizedfinancedefiattacks,web3sec177_sokpreventingtransactionreorderingmanipulations,web3sec176_blockchainimitationgame}. Table~\ref{tab:defi-incidents} lists representative incidents.

\begin{table*}[t]
\centering
\caption{Landmark cross-domain and DeFi incidents.}
\label{tab:defi-incidents}
\scriptsize
\setlength{\tabcolsep}{4.2pt}
\renewcommand{\arraystretch}{1.05}
\setlength{\dashlinedash}{0.65pt}
\setlength{\dashlinegap}{0.95pt}
\setlength{\arrayrulewidth}{0.35pt}
\resizebox{\textwidth}{!}{%
\begin{tabular}{|c|c|c|c|}
\hline
\textbf{Incident} &
\textbf{Approx. loss} &
\textbf{Architectural layer} &
\textbf{Root cause} \\
\hline\hline
The DAO (2016) \cite{web3sec121_makingsmartcontractssmarter,web3sec126_teethergnawingethereumautomaticallyexploit} &
\$60M &
Application &
Fallback logic flaw and reentrancy \\
\hdashline
Parity multisig freeze (2017) \cite{web3sec122_surveyattacksethereumsmartcontracts,web3sec162_proxyhuntingunderstandingcharacterizingproxy} &
\$150M+ &
Application / governance &
Library ownership and upgradeability failure \\
\hdashline
bZx (2020) \cite{web3sec175_attackingdefiecosystemflashloans,web3sec174_sokdecentralizedfinancedefiattacks} &
\$1M &
Application / oracle &
Flash-loan manipulation of AMM spot price \\
\hdashline
Harvest Finance (2020) \cite{web3sec175_attackingdefiecosystemflashloans,web3sec170_safeguardingdefismartcontractsagainst} &
\$34M &
Application / oracle &
Flash-loan price manipulation across stablecoin pools \\
\hdashline
Cream Finance (2021) \cite{web3sec174_sokdecentralizedfinancedefiattacks,web3sec155_sailfishvettingsmartcontractstate} &
\$130M &
Application / lending &
Reentrant lending interaction and collateral accounting flaw \\
\hdashline
Poly Network (2021) \cite{web3sec196_soksecuritycrosschainbridges,web3sec197_smartaxedetectingcrosschainvulnerabilities} &
\$611M &
Cross-domain bridge &
Cross-chain message verification and authorization bypass \\
\hdashline
Wormhole (2022) \cite{web3sec196_soksecuritycrosschainbridges,web3sec187_zkbridgetrustlesscrosschainbridges} &
\$320M &
Cross-domain bridge &
Forged guardian verification and wrapped-asset minting \\
\hdashline
Ronin Bridge (2022) \cite{web3sec196_soksecuritycrosschainbridges} &
\$625M &
Cross-domain bridge &
Validator-key compromise and small committee trust \\
\hdashline
Nomad Bridge (2022) \cite{web3sec196_soksecuritycrosschainbridges,web3sec197_smartaxedetectingcrosschainvulnerabilities} &
\$190M &
Cross-domain bridge &
Misconfigured message root enabling replay-style draining \\
\hdashline
Beanstalk (2022) \cite{web3sec174_sokdecentralizedfinancedefiattacks,web3sec175_attackingdefiecosystemflashloans} &
\$182M &
Application / governance &
Flash-loan governance takeover \\
\hdashline
Mango Markets (2022) \cite{web3sec170_safeguardingdefismartcontractsagainst,web3sec179_uncoveringsecuritypitfallchainlinksoff} &
\$116M &
Oracle domain &
Off-chain exchange manipulation reflected in oracle data \\
\hdashline
Terra / UST collapse (2022) \cite{web3sec174_sokdecentralizedfinancedefiattacks,web3sec200_formalframeworkeconomicsecuritydefi} &
\$40B+ &
Application / stablecoin &
Algorithmic stablecoin death spiral and liquidity run \\
\hdashline
Euler Finance (2023) \cite{web3sec178_clockworkfinanceautomatedanalysiseconomic,web3sec180_houstonrealtimeanomalydetection} &
\$197M &
Application / lending &
Logic flaw bypassing collateral health checks \\
\hdashline
Curve / Vyper pools (2023) \cite{web3sec174_sokdecentralizedfinancedefiattacks,web3sec168_smartcontractdefisecuritytools} &
\$70M+ &
Application / compiler &
Compiler bug enabling reentrancy in deployed pools \\
\hdashline
Bybit / Safe signing flow (2025) \cite{web3incident001_bybithack2025} &
\$1.5B &
Custody / signing infrastructure &
Manipulated cold-to-warm wallet transfer and signing workflow compromise \\
\hdashline
Drift Protocol (2026) \cite{web3incident002_drifthack2026} &
\$280M &
Application / governance &
Durable-nonce social engineering and Security Council control abuse \\
\hdashline
Step Finance (2026) \cite{web3incident003_stepfinancehack2026} &
\$40M &
Application / operational security &
Executive-device compromise enabling illicit treasury access \\
\hline
\end{tabular}%
}
\vspace{2pt}
\begin{flushleft}
\scriptsize
Loss values are approximate and are used only to indicate order of magnitude. The root cause column abstracts each incident into the reusable security pattern discussed in the survey. Recent 2025--2026 rows rely on public incident reports and are not counted as retained academic studies.
\end{flushleft}
\end{table*}

\begin{figure}[t]
  \centering
  \surveyfig{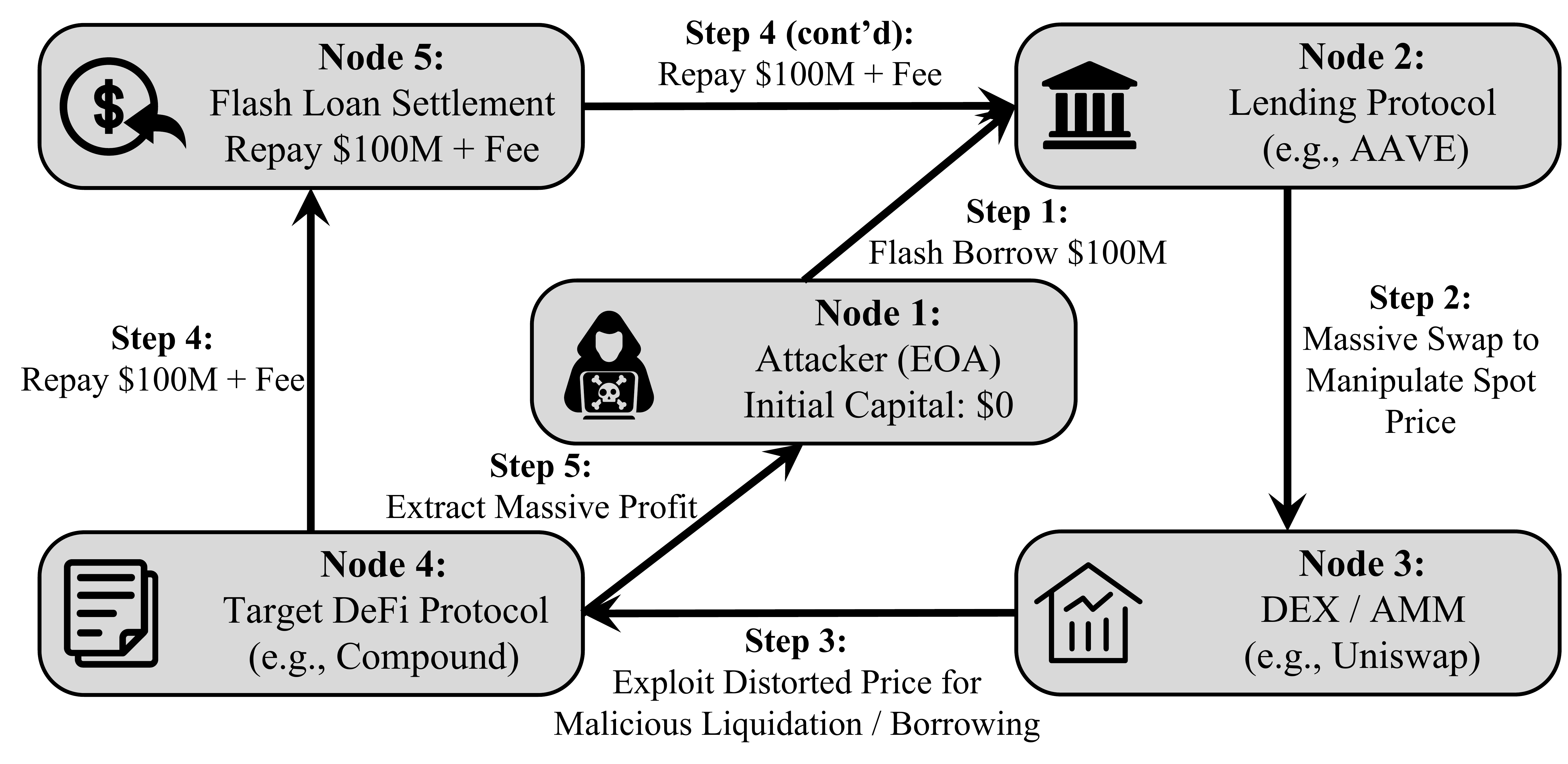}
  \caption{A typical flash-loan-enabled DeFi attack chain.}
  \Description{A schematic showing a flash-loan attack that borrows funds, manipulates AMM reserves, exploits a dependent protocol, and repays the loan in one transaction.}
  \label{fig:flashloan}
\end{figure}

\heading{Flash loans and low-capital-cost market manipulation.}
Flash loans eliminate the capital barrier assumed by traditional finance. They allow an attacker to borrow massive liquidity with no collateral, provided the debt is repaid in the same transaction \cite{web3sec175_attackingdefiecosystemflashloans}. As shown in Figure~\ref{fig:flashloan}, a typical attack borrows funds, distorts AMM reserves, creates a transient price deviation, triggers malicious liquidation or arbitrage in a dependent lending protocol, repays the loan, and keeps the profit. Flash loans can also be used to obtain temporary voting power for governance takeover.

\heading{AMM math errors and token incompatibilities.}
DeFi protocols rely on pricing curves such as $x \times y = k$. Rounding errors and edge-case truncation can create risk-free arbitrage. Token-standard incompatibilities are subtler \cite{web3sec133_tokenscopeautomaticallydetectinginconsistentbehaviors}. Deflationary tokens can desynchronize protocol accounting from actual balances, while ERC-777-style callback hooks can reintroduce control flow into token transfers and create cross-protocol reentrancy. Defenses include constraint-solving-based economic verification with Clockwork Finance \cite{web3sec178_clockworkfinanceautomatedanalysiseconomic}, automated flash-loan payload synthesis with FlashSyn \cite{web3sec171_flashsynflashloanattacksynthesis}, oracle-deviation defenses \cite{web3sec170_safeguardingdefismartcontractsagainst}, and real-time monitoring with HOUSTON and SMARTCAT-style price-manipulation detection \cite{web3sec180_houstonrealtimeanomalydetection,web3sec211_smartcatpricemanipulation}.

\heading{State-dependent risk in composable markets.}
DeFi risk is highly state-dependent. The same contract can move from safe to unsafe when liquidity thins, oracle updates lag, or transaction ordering becomes adversarial. Security is therefore not a static property of bytecode; it is a property of code under a particular market and mempool state.

Audit claims are easier to interpret when they state market assumptions explicitly, and post-deployment monitors can then check whether those assumptions still hold. The goal is to detect when ordinary market behavior invalidates protocol assumptions, not to label every arbitrage as malicious.

\heading{Front-running as an application bug.}
Front-running is often classified as a consensus or mempool problem, but many front-running vulnerabilities are created by application design. If a contract reveals a profitable intent before binding the user to a price, route, or deadline, it invites extraction. If a liquidation or auction mechanism rewards the first visible caller without commit-reveal, threshold encryption, or batch clearing, it transfers value to the participant with the best ordering access. Nyx and related work show that exploitable front-running patterns can be detected at the contract level \cite{web3sec164_nyxdetectingexploitablefrontrunning}. Ordering adversaries belong in the application threat model.

\subsection{Web3 Frontend Threats and Beyond}

As low-level protocols harden, attack surfaces migrate to DApp frontends and governance interfaces. DArcher exposes off-chain view synchronization bugs caused by asynchronous on-chain updates \cite{web3sec153_darcherdetectingchainoffchain}. Proxy upgrade patterns introduce lifecycle risks in ownership transfer and implementation replacement \cite{web3sec162_proxyhuntingunderstandingcharacterizingproxy}. Verification services can be abused so that malicious code masquerades as open-source contracts in block explorers \cite{web3sec163_abusingethereumsmartcontractverification}. At the user-interaction layer, signature phishing can hide real execution semantics behind complex social engineering \cite{web3sec182_phishingwonderlandevaluatinglearningbased}; address poisoning and payload-based transaction phishing show that attackers increasingly exploit wallet display conventions, user memory, and opaque calldata rather than only contract code \cite{web3sec208_addresspoisoning,web3sec215_payloadphishingethereum}. Cryptocurrency technical-support scams remain a direct vector for personal asset theft \cite{web3sec183_conningcryptoconmanendend}.

Frontend security is difficult because wallets compress complex execution into short prompts. Attackers exploit this semantic compression by presenting dangerous authorizations as benign actions. Stronger warnings help only up to the point of fatigue; more durable defenses include transaction simulation with asset-delta summaries, domain-bound approvals, short-lived allowances, hardware-backed confirmation for high-value actions, and human-readable intent formats.

The user interface is part of the security boundary. A wallet that shows an address, a method name, or a raw permission does not necessarily show the state transition the user is authorizing. This mismatch is most dangerous for approvals, permit signatures, upgrade votes, and multisignature operations, where the visible action may be separated from the later asset movement. Safer interfaces bind display, simulation, and signing context: the domain, counterparty, asset delta, deadline, and revocation path are visible before the signature is created.

\heading{Intent-based execution and solver trust.}
The frontend layer is moving from transaction construction to intent expression: users specify outcomes while off-chain solvers choose routes. This can reduce direct MEV exposure, but it moves risk to solver selection, private order flow, settlement, and liquidity routing \cite{web3sec227_analysisintentmarkets}. Cross-chain intents are sharper still: a solver may fill a destination leg before source-chain settlement is final, or depend on bridge liquidity that disappears before reimbursement. Liquidity-exhaustion and proof-carrying intent work show that intent security depends on solver identity, route assumptions, liquidity guarantees, timeouts, and failure compensation being visible to the user or wallet \cite{web3sec228_liquidityexhaustionintents,web3sec229_omniintent}.

Intent systems also change accountability. In a transaction-centric design, responsibility is tied to a concrete call sequence. In an intent-centric design, the critical decisions may be made by solvers, relays, auctions, and settlement services after the user signs. A mature intent protocol produces receipts that connect the authorized outcome to the chosen route, price, solver identity, and failure handling. Without such receipts, users may receive a correct-looking outcome while the system hides who took risk, who extracted value, and who is responsible when cross-chain settlement fails.

\subsection{Summary and Insights}

Application attacks split into two groups. Code-level bugs such as reentrancy, arithmetic errors, and permission mistakes are well covered by scanners and fuzzers, although exploitability evidence remains uneven. Financial-logic attacks depend on market state, oracle timing, governance power, and transaction ordering; these are harder to benchmark because the vulnerable condition may exist only for a few blocks. The missing evaluation layer is economic feasibility: tools need to report the liquidity, gas, ordering, and cross-protocol state required to turn a warning into loss.

\section{Cross-Layer and Cross-Domain Security}
\label{sec:crossdomain}

As Web3 moves toward multi-chain interoperability, off-chain execution, and real-world data integration, single-stack security does not guarantee system robustness. Bridges, rollups, oracles, and payment channels import assumptions from one domain and export consequences to another. If the imported assumption is weaker than the exported consequence, assets may appear protected by one chain while actually depending on a smaller validator set, an off-chain server, a centralized sequencer, or a delayed price feed. Table~\ref{tab:crossdomain-failures} lists the recurring failure modes.

These modes are not mutually exclusive. A bridge exploit may combine validation asymmetry with semantic mismatch; a rollup failure may combine data unavailability with sequencer censorship; an oracle attack may combine timing lag with off-chain market manipulation. Cross-domain risk is therefore rarely reducible to one vulnerable contract.

\begin{table*}[t]
\centering
\caption{Common cross-domain failure modes.}
\label{tab:crossdomain-failures}
\scriptsize
\setlength{\tabcolsep}{4.6pt}
\renewcommand{\arraystretch}{1.05}
\setlength{\dashlinedash}{0.65pt}
\setlength{\dashlinegap}{0.95pt}
\setlength{\arrayrulewidth}{0.35pt}
\resizebox{\textwidth}{!}{%
\begin{tabular}{|c|c|c|}
\hline
\textbf{Failure mode} &
\textbf{Typical setting} &
\textbf{Security consequence} \\
\hline\hline
Validation asymmetry \cite{web3sec196_soksecuritycrosschainbridges,web3sec197_smartaxedetectingcrosschainvulnerabilities} &
Bridges, rollups, light clients &
A weak verifier controls high-value assets in another domain \\
\hdashline
Finality mismatch \cite{web3sec186_speculartowardssecuretrustminimized,web3sec204_bscfinalitygadget} &
Cross-chain bridges, L2-to-L1 settlement &
A destination domain accepts state before the source domain is economically final \\
\hdashline
Timing mismatch \cite{web3sec193_sleepychannelsbidirectionalpayment,web3sec194_breakingfixingutxobasedvirtual} &
Channels, optimistic rollups, oracles &
Honest responses arrive after the economically relevant deadline \\
\hdashline
Semantic mismatch \cite{web3sec186_speculartowardssecuretrustminimized,web3sec210_polygonzkevmfreever} &
Cross-chain messages, VM variants, fraud proofs &
Two domains interpret the same state transition differently \\
\hdashline
Replay or domain-separation failure \cite{web3sec196_soksecuritycrosschainbridges,web3sec197_smartaxedetectingcrosschainvulnerabilities} &
Bridges, signature-based message passing &
A valid message in one context is reused in another context \\
\hdashline
Data-availability gap \cite{web3sec185_arbitrumscalableprivatesmartcontracts,web3sec186_speculartowardssecuretrustminimized} &
Rollups, bridges, fraud proofs &
Invalid state cannot be challenged because evidence is missing \\
\hdashline
Sequencer or relayer centralization \cite{web3sec185_arbitrumscalableprivatesmartcontracts,web3sec187_zkbridgetrustlesscrosschainbridges} &
Rollups, externally verified bridges &
A privileged operator can censor, delay, reorder, or selectively relay messages \\
\hdashline
Oracle-source correlation \cite{web3sec170_safeguardingdefismartcontractsagainst,web3sec179_uncoveringsecuritypitfallchainlinksoff} &
DeFi lending, derivatives, liquidation protocols &
Nominally independent feeds fail together due to shared data sources or infrastructure \\
\hdashline
Accounting or solvency mismatch \cite{web3sec196_soksecuritycrosschainbridges,web3sec051_provisionsprivacypreservingproofssolvency} &
Lock-and-mint bridges, wrapped assets, stablecoins &
Synthetic claims exceed real collateral or reserves \\
\hdashline
Liquidity and routing mismatch \cite{web3sec195_atomicityefficiencyblockchainpaymentnetworks,web3sec189_universalatomicswapssecureexchange} &
Cross-chain swaps, payment channels, cross-rollup routing &
One leg of a multi-domain transaction executes while another fails or becomes illiquid \\
\hdashline
Governance mismatch \cite{web3sec196_soksecuritycrosschainbridges,web3sec162_proxyhuntingunderstandingcharacterizingproxy} &
Upgradeable bridges, oracle networks, L2 admin keys &
Emergency control in one domain becomes hidden control over another \\
\hdashline
Incentive mismatch \cite{web3sec181_deficomposabilitymevnoninterference,web3sec203_bunnyfinderethereumconsensus} &
MEV markets, bridge validators, oracle committees &
Rational participants profit by delaying, censoring, or deviating from the expected role \\
\hdashline
Shared-security overload \cite{web3sec225_robustrestakingnetworks,web3sec226_financialdynamicsrestaking} &
Restaking, AVSs, liquid-restaking derivatives &
One service's slashing or liquidity shock propagates through reused collateral and shared operators \\
\hline
\end{tabular}%
}
\vspace{2pt}
\begin{flushleft}
\scriptsize
The modes are not mutually exclusive. Real incidents usually combine several mismatches, which is why cross-domain failures often escape single-layer audits.
\end{flushleft}
\end{table*}

\subsection{Chain-to-Chain: Cross-Chain Bridge Vulnerabilities}

Transferring assets and state across networks with independent consensus mechanisms is necessary to overcome blockchain data silos. Early interoperability designs used hash time-locked contracts (HTLCs) and trust-minimized multi-hop exchange mechanisms \cite{web3sec184_xclaimdecentralizedinteroperablecryptocurrencybacked,web3sec188_hyperserviceinteroperabilityprogrammabilityacrossheterogeneous,web3sec189_universalatomicswapssecureexchange}. Due to fragmented liquidity and interaction latency, however, current ecosystems rely heavily on cross-chain bridges that proxy state synchronization.

\begin{figure}[t]
  \centering
  \surveyfig{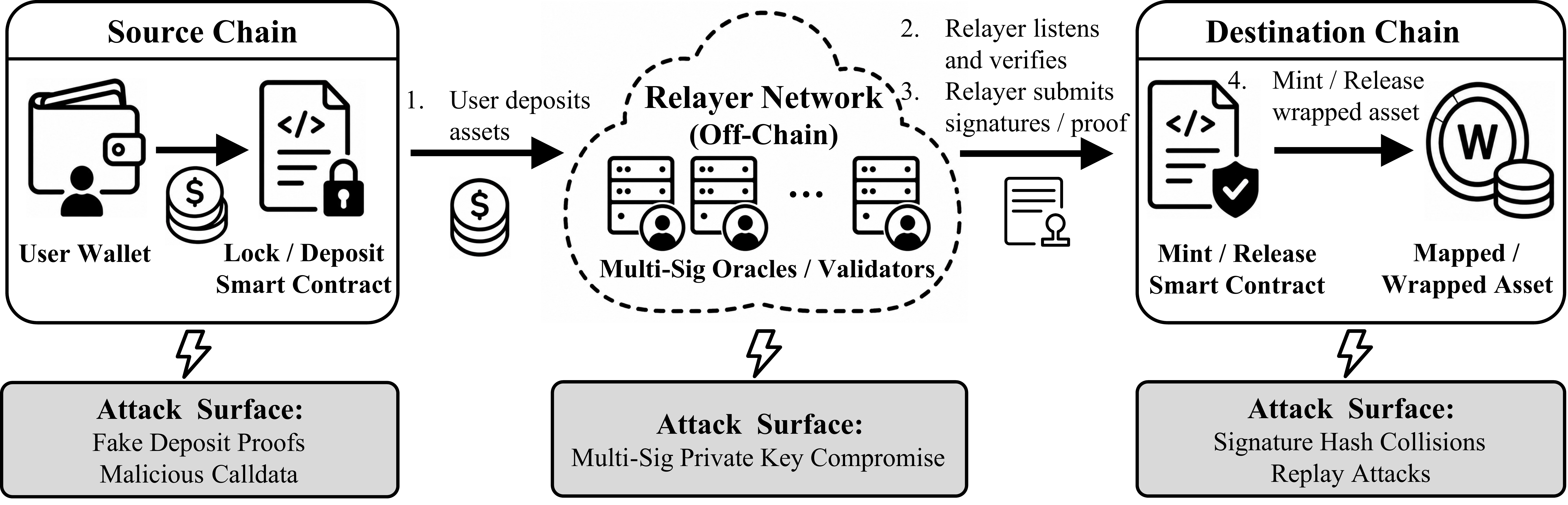}
  \caption{Trust boundaries in a typical cross-chain bridge.}
  \Description{A schematic showing source-chain contracts, relayers or oracle networks, destination-chain contracts, and the verification path across bridge trust domains.}
  \label{fig:bridge}
\end{figure}

A bridge projects the finality assumption of a source chain onto a potentially heterogeneous destination chain. As Figure~\ref{fig:bridge} shows, a typical bridge spans three trust domains: source-chain lock or burn contracts, off-chain relayers or oracle networks, and destination-chain mint or release contracts. Recent SoK work on bridge security, SmartAxe's cross-chain static analysis, and Alba's scalable bridge design reveal the structural fragility and performance pressure of this hybrid architecture \cite{web3sec196_soksecuritycrosschainbridges,web3sec197_smartaxedetectingcrosschainvulnerabilities,web3sec213_albabridges}.

\heading{Polarized attack surfaces.}
Externally verified bridges concentrate trust in off-chain relayers or multisignature validator sets. If attackers steal a majority of validator keys through phishing or server compromise, as in several major incidents, they can bypass the source-chain lock logic and issue valid minting instructions on the destination chain. Light-client-verified bridges reduce this assumption but introduce cross-language implementation risks across Solidity, Rust, Go, and cryptographic libraries. Bit-level differences in hash-tree parsing or signature verification can enable signature-collision-style attacks against destination-chain verification contracts.

To remove privileged relayer assumptions, research is shifting toward trustless ZK bridges. zkBridge generates off-chain ZK proofs of source-chain headers and consensus evolution, then verifies them on the destination chain with low gas overhead, providing cryptographic consistency across domains \cite{web3sec187_zkbridgetrustlesscrosschainbridges}.

\heading{Liquidity, accounting, and failure containment.}
Bridge security is also accounting security. Lock-and-mint bridges create synthetic assets whose value depends on locked collateral; if wrapped assets exceed collateral, the failure becomes a bank run. Liquidity networks and canonical bridges still need explicit route, operator, and liquidity-risk allocation. The measurable fields are blast radius, abnormal-flow threshold, slow-path trigger, and maximum loss before intervention.

This containment view changes the design goal. A bridge optimized only for speed and liquidity is effectively selling insurance without reserves. Rate limits, delayed settlement for abnormal flows, independent monitoring, and clear pause or exit rules may reduce convenience, but they turn catastrophic compromise into a loss that users and protocols can reason about.

\heading{Upgradeability and social trust.}
Many bridges retain upgrade keys for emergency response, but upgradeability changes the trust model. Users may believe they rely on source-chain consensus while actually relying on administration. A transparent bridge exposes who can change the system, how slowly changes take effect, and how users can exit before changes bind them.

\subsection{L2-to-L1: Rollup and Off-Chain Scaling Security}

Layer-2 scaling offloads execution from Layer-1 (L1) to overcome throughput constraints. This separation of execution from settlement introduces complex cross-domain games between the execution layer and arbitration layer.

\heading{Timing and topological fragility in channels.}
Payment and state-channel networks, including Lightning-style systems, assume that users can submit penalty or justice transactions to L1 within a challenge window after malicious closure \cite{web3sec190_atomicmultichannelupdatesconstant,web3sec191_perunvirtualpaymenthubsover,web3sec192_generalstatechannelnetworks,web3sec193_sleepychannelsbidirectionalpayment}. If attackers trigger BDoS or targeted congestion at critical times \cite{web3sec194_breakingfixingutxobasedvirtual}, or exploit channel-network topology to exhaust liquidity \cite{web3sec195_atomicityefficiencyblockchainpaymentnetworks}, honest dispute transactions may fail to be included on L1. Congestion thus strips channel networks of their economic punishment mechanism.

\heading{Asymmetric verification in rollups.}
Optimistic and ZK rollups typically use L2 sequencing and execution while relying on L1 for data availability and state verification. This design introduces security degradation \cite{web3sec185_arbitrumscalableprivatesmartcontracts}: the user receives a faster interface, but the safety claim now depends on sequencer behavior, semantic agreement between execution environments, and the correctness of the proof or dispute path. Specular shows that small differences between an L2 execution engine and the L1 arbitration semantics used in fraud proofs can allow malicious sequencers to construct valid-looking but false state transitions \cite{web3sec186_speculartowardssecuretrustminimized}. For ZK rollups, automated vetting of Polygon zkEVM shows that proving-system implementations need the same adversarial testing discipline as virtual machines \cite{web3sec210_polygonzkevmfreever}. Escape hatches for forced withdrawal must also be formally verified; otherwise, catastrophic L2 failure can permanently lock user funds.

\heading{Data availability and the right to exit.}
The strongest rollup claim is user exit under operator failure. That claim depends on data availability: without transaction data, users cannot produce fraud proofs, verify ZK transitions, or compute exit balances. Sequencer decentralization alone is insufficient; evaluation records whether users can exit without sequencer cooperation, whether challengers can act during congestion, and whether L1 contracts can adjudicate disputes without ambiguous L2 semantics.

The right-to-exit property needs stress tests rather than diagram-level descriptions. Forced-inclusion lists, challenge windows, data-availability committees, proof-verifier upgrades, and emergency exits must be evaluated together because failure in any one component can turn a temporary L2 outage into permanent asset lockup.

\heading{Cross-rollup composition.}
Cross-rollup composition adds another timing layer. A trade may begin on one rollup, depend on liquidity on another, and settle through L1, crossing several finality clocks. Routing quality depends on settlement assumptions and credit risk as well as the quoted price.

\subsection{Off-Chain to On-Chain: The Oracle Problem}

Smart contracts are closed deterministic state machines and cannot natively observe asset prices, weather, flights, or other real-world states. Oracles import Web2 data into Web3 and thereby build a projection from nondeterministic real-world systems into deterministic cryptographic protocols. Oracle failures exploit incorrect data, time lags, and mechanism asymmetry.

\heading{Data-source pollution and decentralized-consensus weaknesses.}
Attackers need not break a blockchain if they can poison data providers using Web2 attacks such as BGP hijacking, DNS spoofing, or API-key theft. Decentralized oracle networks try to reduce single points of failure through aggregation, but their off-chain aggregation protocols may contain game-theoretic weaknesses. Recent analysis of Chainlink's off-chain reporting protocol shows consensus fragility under certain Byzantine attacks, allowing a minority of malicious nodes to bias aggregated prices \cite{web3sec179_uncoveringsecuritypitfallchainlinksoff}.

\heading{Cross-domain arbitrage.}
Cross-domain arbitrage is among the most destructive DeFi attack paradigms. Attackers exploit the lag between L1 block inclusion and oracle-price updates. A typical attack uses large trades in a low-liquidity centralized exchange to crash an asset's spot price, then, before the oracle synchronizes the drop, triggers malicious liquidation in an on-chain lending protocol \cite{web3sec170_safeguardingdefismartcontractsagainst}. This converts ordinary off-chain market volatility into an on-chain deterministic exploit.

\heading{Oracle design as adversarial statistics.}
Oracle aggregation is both a decentralization problem and a statistical problem under adversarial sampling. Medians help only when feeds are independent, timely, and hard to manipulate together. Shared exchange data, cloud infrastructure, or reporting committees can defeat nominal decentralization. Time-weighted prices reduce spikes but add lag; fast updates reduce lag but become reactive to manipulated markets. Robust applications propagate oracle confidence into collateral requirements, liquidation speed, and borrow caps instead of hiding uncertainty behind one price.

Oracle assumptions depend on market microstructure. A feed for a highly liquid asset can use different windows, venues, and deviation thresholds than a feed for a thin governance token. Treating all prices as equally trustworthy encourages protocols to overreact in exactly the markets where manipulation is cheapest.

\subsection{Cross-Layer Synergistic Attacks}

In complex Web3 systems, high-capability adversaries often combine layers rather than attack a single subsystem. Cross-layer attacks violate the assumptions that upper-layer mechanisms make about lower-layer synchrony, connectivity, or ordering.

\heading{Topological interference with consensus and privacy.}
In payment-channel networks, routing privacy conflicts with high-throughput routing requirements. Attackers can deploy network-layer probing nodes to infer routing graphs and undermine application-layer cryptographic commitments, damaging channel balance privacy \cite{web3sec033_highthroughputcryptocurrencyroutingpayment,web3sec034_privacyutilitytradeoffsroutingcryptocurrency,web3sec035_madhtlcbecausehtlccrazy,web3sec036_hehtlcrevisitingincentiveshtlc,web3sec037_settlingpaymentsfastprivateefficient}. For consensus, BGP hijacking that delays a major mining pool, or eclipse attacks that isolate validators, can mathematically lower the hash power required for successful double spending.

\heading{Consensus timing against application logic.}
Smart contracts often assume that legitimate transactions will be included within expected block-time bounds. If attackers congest the base chain using BDoS or bribe validators to censor selected interactions, time-dependent application defenses such as mutexes, liquidation bots, timelocks, or fraud proofs may fail. Recent mechanism-design work shows that unbounded DeFi composability amplifies the systemic MEV risk caused by consensus-layer transaction ordering and can break application-level noninterference assumptions \cite{web3sec181_deficomposabilitymevnoninterference}.

\subsection{Defensive Principles for Cross-Domain Systems}

Cross-domain defenses fail when imported claims have no explicit value cap, timeliness bound, override path, or exit route. Direct verification reduces relayer trust, but many systems still rely on small committees without publishing the value at risk. Time-dependent mechanisms face a similar gap: delays are often documented as performance parameters rather than priced as security risk. Rate limits, exits, and slashing boundaries are the mechanisms that keep such failures local, especially in restaking systems where a local AVS dispute can create pressure on the base chain.

A practical coding field is the \emph{security budget} of an imported claim. The budget records maximum value released, finality depth assumed, actors trusted, monitoring delay tolerated, and emergency action available if the claim is disputed. A bridge message, oracle price, rollup state root, or intent receipt can then carry a risk envelope rather than only data. This makes stale prices, abnormal bridge flows, congestion, and high-value transfers measurable conditions for stricter verification.

Upgrade governance is another imported assumption. Many systems describe emergency powers in documentation rather than in the security model. The relevant fields are the committee that can pause a bridge, upgrade a verifier, rotate oracle members, or modify an L2 exit path; the timelock before activation; the veto or exit window; and the public evidence available before the change takes effect.

\section{Open Challenges and Future Directions}
\label{sec:future}

The open problems below are phrased as unresolved measurement or modeling questions. They focus on information that current papers often omit: adversarial traces, migration state, solver receipts, slashing exposure, disclosure governance, hardware trust, and time-to-action.

\subsection{The Double-Edged Sword of LLMs in Web3 Security}

AI-security surveys, GNN scanners, and hybrid systems such as GPTScan show that learned representations can find financial-logic vulnerabilities beyond traditional symbolic execution \cite{web3sec198_sokaipoweredsecurityanalysis,web3sec161_smartercontractsdetectingvulnerabilitiessmart,web3sec166_scvhuntersmartcontractvulnerabilitydetection,web3sec169_gptscandetectinglogicvulnerabilitiessmart}. The weak point is evaluation. Existing benchmarks rarely include private order flow, failed exploit attempts, same-block competition, or solver behavior, even though these factors decide whether an AI-generated attack plan is feasible on-chain. FORGE-like corpora are an early response, but they still leave the market-state problem open \cite{web3sec199_forgellmdrivenframeworklarge}.

The unresolved question is which model outputs can be converted into evidence. Candidate invariants, suspicious call sequences, or natural-language exploit explanations need translation into symbolic queries, fuzz targets, transaction replays, or human-checkable proof obligations. Defensive use has the same requirement: a model flag for oracle manipulation or governance takeover becomes actionable only when tied to predicates such as price deviation, borrowed capital, affected collateral, privilege change, or settlement status.

\subsection{Post-Quantum Migration of Blockchain Cryptography}

Web3 authentication, consensus validation, and many proof systems rely on assumptions threatened by large-scale quantum computers. Lattice- and hash-based designs, including MatRiCT and MatRiCT+ \cite{web3sec058_matrictefficientscalablepostquantum,web3sec059_matrictplusmoreefficientpostquantum}, show that post-quantum privacy is possible, while formal methods can help check new arithmetic \cite{web3sec064_certifyingzeroknowledgecircuitsrefinement}. The unresolved migration questions are concrete: how to handle dormant accounts whose owners never rotate keys, old signatures that remain valid in bridge or light-client verification, larger signatures that change propagation cost, and contracts that hard-code curve-specific assumptions.

\subsection{Compositional Security in Complex DeFi Ecosystems}

Composability creates vulnerabilities that isolated contract analysis misses. Economic abstraction languages for nested DeFi systems \cite{web3sec200_formalframeworkeconomicsecuritydefi} need to connect game theory, control theory, and noninterference reasoning under malicious MEV \cite{web3sec181_deficomposabilitymevnoninterference}. The open modeling question is how to represent both atomic execution and continuously changing prices, liquidity, and ordering conditions. Auditors need counterfactual answers: how much liquidity movement breaks solvency, what ordering advantage makes an exploit profitable, and which dependency dominates risk.

Reusable interface specifications are still missing. A lending market, AMM, oracle, and bridge expose function signatures, but usually not tolerated price lag, liquidity depth, finality delay, or failure behavior. Without those fields, composition risk is reconstructed after an incident from source code, documentation, and governance history.

\subsection{Restaking Risk Containment}

Restaking turns validator collateral into shared security, but reused collateral can turn one service's bug into system-wide loss \cite{web3sec225_robustrestakingnetworks,web3sec226_financialdynamicsrestaking}. The unresolved modeling problem is AVS blast radius. A usable model needs correlated operator sets, correlated slashing events, liquid-restaking leverage, governance intervention probability, and the maximum loss that can be imposed on validators who did not directly participate in the faulty service.

\subsection{Intent-Centric Security}

Intent-based execution shifts verification from calldata correctness to market behavior: who fills the intent, what route they choose, and who absorbs failure across chains \cite{web3sec227_analysisintentmarkets,web3sec228_liquidityexhaustionintents}. The concrete open question is the content of a solver receipt. At minimum, a receipt needs the authorized outcome, solver identity or accountability handle, route, price bound, liquidity source, timeout, bridge or settlement dependency, and failure compensation rule. Proof-carrying intent work begins to formalize this record, but deployment practice is still immature \cite{web3sec229_omniintent}.

\subsection{Compliance-Compatible Privacy and Decentralization}

Absolute on-chain privacy protects legitimate confidentiality but conflicts with tracing requirements. Selective disclosure is a middle path: users remain private during ordinary use but can generate proofs of lawful provenance, clean funds, or solvency for authorized review \cite{web3sec051_provisionsprivacypreservingproofssolvency}. The open problem is disclosure governance: who authorizes a request, what scope is revealed, how abuse is detected, and how an honest user proves compliance without creating a reusable surveillance key.

\subsection{Hardware-Assisted Web3 Security}

Cross-chain interoperability and Layer-2 execution make pure software verification expensive. TEEs can harden bridge validators, protect rollup sequencers, or provide auditable ordering traces, and recent rollback-resilience work shows how hardware roots of trust can support accountability when paired with protocol safeguards \cite{web3sec217_pallasaegisrollback}. The missing evidence concerns the residual hardware trust: vendor compromise, attestation freshness, rollback detection, side-channel exposure, and whether users can exit when the hardware root is disputed.

\subsection{Reproducibility, Datasets, and Evaluation Standards}

Blockchain security research often depends on historical, private, proprietary, or rapidly changing data. Current datasets preserve exploit transactions more often than execution context. DeFi feasibility depends on surrounding state, MEV data misses failed bundles and private order flow, and bridge or oracle incidents may hinge on off-chain logs and governance actions. Evaluations need separate fields for detection, explanation, and mitigation, with explicit assumptions, replayability, economic feasibility, and response time.

A useful benchmark also records what it cannot observe. Missing private order flow, unavailable off-chain committee logs, incomplete mempool data, and post-incident governance actions can change the interpretation of an attack. Reporting these blind spots makes negative results more credible and helps compare tools that operate at different stages: pre-deployment audit, runtime monitoring, incident forensics, and automated mitigation.

Finally, evaluation needs a time-to-action field. In blockchains, a true positive after finality may be useful for forensics and useless for prevention. A fuzzer finding a vulnerable path matters only if the required market state is reachable. A runtime monitor is valuable only if it can trigger before settlement or route users to an exit. Standard reports can include detection time, response time, required privileges, capital assumptions, replay scripts, and whether the mitigation still works under congestion.

\section{Conclusion}
\label{sec:conclusion}

This paper reviews blockchain attacks and defenses, with an emphasis on the shift from isolated single-component vulnerabilities to complex cross-domain and composable exploits. We categorize the threat landscape across a layered architecture—encompassing network, cryptographic, consensus, and application layers—and provide insights into how security assumptions propagate across trust boundaries. We also summarize representative mitigation strategies and compare them by their operational trade-offs, timing requirements, and residual risks. Finally, we examine real-world failures in cross-domain infrastructures such as bridges, rollups, and decentralized finance (DeFi) ecosystems. By reviewing the evolution of Web3 security and analyzing practical exploits, we summarize the research trend toward compositional security, provide an outlook for future decentralized architectures, and discuss the open challenges that need to be addressed.

\bibliographystyle{ACM-Reference-Format}
\bibliography{ref}

\end{document}